\documentclass[%
 reprint,
 amsmath,amssymb,
 aps,
 nofootinbib,
]{revtex4-2}

\usepackage[breaklinks=true]{hyperref}
\usepackage{graphicx}
\usepackage{dcolumn}
\usepackage{bm}

\begin{document}

\preprint{APS/123-QED}

\title{Probing a scale dependent gravitational slip with galaxy strong lensing systems
}

\author{Sacha Guerrini}
 \email{sacha.guerrini@polytechnique.edu}
 \affiliation{Ecole Polytechnique, Palaiseau, F-91128, France}
\author{Edvard Mörtsell}%
 \email{edvard@fysik.su.se}
\affiliation{%
 The Oskar Klein Centre, Department of Physics, Stockholm University,\\
 Albanova University Center, Stockholm, SE-106 91, Sweden
}%

\date{\today}

\begin{abstract}
Observations of galaxy-scale strong gravitational lensing systems enable unique tests of departures from general relativity at the kpc-Mpc scale. In this work, the gravitational slip parameter $\gamma_{\rm PN}$, measuring the amplitude of a hypothetical fifth force, is constrained using 130 elliptical galaxy lens systems. We implement a lens model with a power-law total mass density and a deprojected De Vaucouleurs luminosity density, favored over a power-law luminosity density. To break the degeneracy between the lens velocity anisotropy, $\beta$, and the gravitational slip, we introduce a new prior on the velocity anisotropy based on recent dynamical data. For a constant gravitational slip, we find $\gamma_{\rm PN}=0.90^{+0.18}_{-0.14}$ in agreement with general relativity at the 68\% confidence level. Introducing a Compton wavelength $\lambda_g$, effectively screening the fifth force at small and large scales, the best fit is obtained for $\lambda_g \sim 0.2$ Mpc and $\gamma_{\rm PN} = 0.77^{+0.25}_{-0.14}$. A local minimum is found at $\lambda_g \sim 100$ Mpc and $\gamma_{\rm PN}=0.56^{0.45}_{-0.35}$.
We conclude that there is no evidence in the data for a significant departure from general relativity and that using accurate assumptions and having good constraints on the lens galaxy model is key to ensure reliable constraints on the gravitational slip.\\ 

\end{abstract}

\maketitle


\section{Introduction}

Together with quantum field theory, Einstein's theory of general relativity (GR) is a cornerstone of modern physics. Those two theories yield a description of the history of the Universe from a fraction of a second after the Big Bang to today, in what's called the cosmological concordance model $\Lambda$CDM \cite{blanchardEuclidPreparationVII2020}. The latter model is not fully understood however. In particular, the accelerated cosmic expansion remains one of the most puzzling questions in cosmology and in physics in general \cite{weinbergCosmologicalConstantProblem1989}. It may be formally understood as a cosmological constant added to Einstein equations expressing the link between space-time curvature and the stress-energy tensor $T_{\mu \nu}$. The required cosmological constant is very small and presents a discrepancy of $\gtrsim 60$ orders of magnitude with theoretical estimates, refered to as the \emph{cosmological constant problem} \cite{joyceCosmologicalStandardModel2015}.\\

Another perspective for understanding cosmic acceleration is to modify Einstein's theory of gravity \cite{shankaranarayananModifiedTheoriesGravity2022}. So far, GR has been confirmed in all experiments, especially at the Solar System scale \cite{schlammingerTestEquivalencePrinciple2008, shapiroFourthTestGeneral1964, poundApparentWeightPhotons1960} but the true gravity theory might deviate from GR at cosmological scales. Therefore, determining whether dark energy or modified gravity (MG) drives cosmic expansion can potentially be adressed with a test of GR at cosmological scales. Many MG theories can be embedded in a phenomenological description \cite{koyamaCosmologicalTestsModified2016}, allowing for measurements of general departures from GR. The validity of GR can be tested by constraining the gravitational slip parameter $\gamma_{\rm PN}$ \cite{thorneTheoreticalFrameworksTesting1971}, which describes how much space curvature is provided by the unit rest mass of objects. In addition, screening mechanisms appear naturally in many MG theories and restore GR on small and large scales \cite{joyceCosmologicalStandardModel2015}. \\

Several cosmological probes allow tests of GR under screening. Among them, strong gravitational lensing (SGL) occurs due to the curving of space-time induced by mass. Strong lensing more precisely refers to the formation of multiple source images by a lens mass located close to the line of sight towards the source. In recent years, great efforts have been put into estimating cosmological parameters \cite{caoCOSMOLOGYSTRONGLENSINGSYSTEMS2015, amanteTestingDarkEnergy2020}, measuring the Hubble constant $H_0$ \cite{birrerH0LiCOWIXCosmographic2019, wongH0LiCOWXIIICent2020}, the cosmic curvature \cite{liuTestingCosmicCurvature2020} and the distribution of matter in massive galaxies acting as lenses \cite{caoLimitsPowerlawMass2016, chenAssessingEffectLens2019}. Provided reasonable prior assumptions and appropriate descriptions of the internal structure of lensing galaxies, it is possible to constrain the gravitational slip $\gamma_{\rm PN}$ using SGL \cite{caoTESTPARAMETRIZEDPOSTNEWTONIAN2017, weiDirectEstimatePostNewtonian2022, liuGalaxyscaleTestGeneral2022, lianDirectTestsGeneral2022}. Recent publications introduced a phenomenological screening model as a step discontinuity in $\gamma_{\rm PN}$ at a scale $r_V$ \cite{jyotiCosmicTimeSlip2019, adiProbingGravitationalSlip2021, lianDirectTestsGeneral2022}. The obtained constraint in Ref. \onlinecite{jyotiCosmicTimeSlip2019} is $|\gamma_{\rm PN} -1| \leq 0.2 \times (r_V/100\text{ kpc})$ with $r_V = 10 - 200$ kpc using two gravitationally lensed quasars time-delay measurements. Fast radio burst time-delay simulations \cite{adiProbingGravitationalSlip2021}, predict constraints $|\gamma_{\rm PN}-1| \leq 0.04 \times (r_V/100 \text{ kpc}) \times [N/10]^{-1/2}$ where $N$ is the sample size. Ten events alone could place constraints at a level of $10\%$ in the range $r_V=10-300$ kpc.

In this work, we take advantage of a recently compiled sample of 130 SGL systems \cite{chenAssessingEffectLens2019} to investigate a gravitational slip under screening effects. Here, we assume that only massless photons will be affected by the fifth force i.e. only the longitudinal potential, $\Psi$, varies. This is a common assumption \cite{chenAssessingEffectLens2019, lianDirectTestsGeneral2022} motivated by the fact that we only probe the difference between massive and massless particles. We introduce a phenomenological description of screening at small and large scales respectively, parameterised by the Vainshtein radius, $r_V$, and the Compton wavelength of the theory, $\lambda_g$. The combination of lensing and stellar kinematics data is used to constrain possible discrepancies in the gravitational effects on massless (photons) and massive (stars, gas, ...) particles. We introduce a deprojected De Vaucouleurs luminosity density to be compared with the commonly used power-law luminosity profile. We assess the influence of the lens mass model on our estimation of the gravitational slip and finally study the degeneracy between the gravitational slip and the Compton wavelength of the theory for $\lambda_g = 1 \text{pc} - 100\, \text{Gpc}$.

This paper is organised as follows: in Section \ref{seq: Methodology}, we introduce the model used to evaluate the velocity dispersion of lensing galaxies and our phenomenological screening description. We further introduce our SGL sample, the cosmological model as well as the model parameters for which we perform a Markov Chain Monte Carlo (MCMC) analysis. In Section \ref{seq: results and discussion}, we present and discuss our results. The case without screening is first used to asses the influence of the lens mass model on the fit before studying the degeneracy between the Compton wavelength and the gravitational slip. Conclusions are summarised in Section \ref{seq: conclusion}.\\

\section{Methodology}\label{seq: Methodology}

\subsection{The Model}

\subsubsection{The general framework}

The general idea is to measure the mass enclosed inside the Einstein radius of the lens using both massless photons and massive stars as probes of the gravitational potential. Besides the imaging data of the SGL, spectroscopic data of the system is needed to measure the velocity dispersion of the lens galaxy. The comparison of the projected gravitational and dynamical masses ($M_{\rm grav}$ and $M_{\rm dyn}$ respectively) provides a promising test of GR at the galactic scales.\\

From the theory of gravitational lensing, the gravitational mass is $M_{\rm grav} = \Sigma_{\rm cr} \pi R_{E, \text{GR}}^2$ \cite{schneiderGravitationalLensingStrong2006} in GR where $R_{E, \text{GR}} = \theta_{E, \text{GR}} D_l$ is the Einstein radius wherein $\theta_{E, \text{GR}}$ is the Einstein angle and $D_l$ the angular distance between the observer and the lens. The critical surface density is defined by,

\begin{eqnarray}
    \Sigma_{\rm cr} = \frac{c^2}{4\pi G}\frac{D_s}{D_l D_{ls}},
\end{eqnarray}
where $D_s$ and $D_{ls}$ are the angular distances between the observer and the source and between the lens and the source respectively.\\

A mass distribution model of the lens galaxy $[\rho(r), \nu(r), \beta]$ is required to compute the velocity dispersion in the lens galaxy and the dynamical mass $M_{\rm dyn}$. $\rho$ is the total mass density, $\nu$ the luminosity density of stars and $\beta$ the anisotropy of the velocity dispersion assumed to be constant in this work. Assuming spherical symmetry, the Jeans equation \cite{binneyGalacticDynamicsSecond2008} is given by

\begin{eqnarray}\label{eq: Jeans}
    \frac{d}{dr}[\nu(r) \sigma_r^2]+2\frac{\beta}{r}\nu(r)\sigma_r^2=-\nu(r)\frac{d\Phi}{dr},
\end{eqnarray}
where the gravitational potential is given by

\begin{eqnarray}
    \frac{d\Phi}{dr} = \frac{GM(r)}{r^2},
\end{eqnarray}
where $M(r)$ denotes the mass enclosed inside a sphere of radius $r$. After integration,

\begin{eqnarray}\label{eq: sigma2_r}
    \sigma_r^2(r) = \frac{G\int_r^{\infty}dr'\nu(r')r'^{2\beta-2}M(r')}{r^{2\beta}\nu(r)}.
\end{eqnarray}

In an observational context, we do not measure $\sigma^2_r$ but rather the luminosity-weighted average along the line of sight (los) and over the effective spectroscopic aperture $R_A$ \cite{chenAssessingEffectLens2019}. This can be expressed mathematically,

\begin{align}\label{eq: observed VD}
    \sigma^2_{\parallel}(\leq R_A) = \frac{\int_0^{R_A}dR 2\pi R \int_{-\infty}^{\infty}dZ \sigma^2_{los}\nu(r)}{\int_0^{R_A}dR 2\pi R \int_{-\infty}^{\infty}dZ\nu(r)},
\end{align}
where $\sigma_{\rm los}$ is the velocity dispersion along the line of sight,

\begin{align}
    \sigma_{\rm los}^2 = (\sigma_r \cos\theta)^2+(\sigma_t \sin\theta)^2,
\end{align}
where $\sigma_t$ is the tangential velocity dispersion, $\sigma_r$ the radial velocity dispersion and $\theta$ the angle between the line of sight and the radial direction. Note that $\sigma_r^2$ contains $M_{\rm dyn}$ since we use the equality $M_{\rm grav}=M_{\rm dyn}$ to fix the normalisation constant of the density $\rho$.

\subsubsection{Lens mass models}

In this work we use the following lens mass model:

\begin{eqnarray}
    \left\{
    \begin{array}{ccl}
        \rho(r) & = & \rho_0 \left( \frac{r}{r_0}\right)^{-\gamma},  \\
        \nu(r) & = & \nu_0 \left( \frac{r}{a}\right)^{-\delta} \exp \left(-\left( \frac{r}{a}\right)^{1/4} \right),\\
        \beta & = & 1-\frac{\sigma_t^2}{\sigma_r^2},
    \end{array}
\right.
\end{eqnarray}
where $\rho$ follows a commonly used power-law distribution \cite{chenAssessingEffectLens2019, lianDirectTestsGeneral2022, caoTESTPARAMETRIZEDPOSTNEWTONIAN2017} and $\nu$ a deprojected De Vaucouleurs density profile \cite{mellierDeprojectionVaucouleursExp1987} where $a=R_{\rm eff}/b^4$ with $b=7.66925$ and $\delta=0.8556$. It will be compared to the commonly used power-law $\nu_{\rm pl}(r) = \nu_0(r/r_0)^{-\delta}$. The latter is convenient since the velocity dispersion can be expressed analytically \cite{chenAssessingEffectLens2019}. The case of the De Vaucouleurs deprojected luminosity density requires numerical integration:

{\small
\begin{align}\label{eq: los vd dv}
\begin{split}
    \sigma_{\parallel}^2(\leq R_A) = \frac{2c^2}{\sqrt{\pi}} \frac{D_s}{D_{ls}}&\theta_{E, \text{GR}} \frac{\Gamma(\gamma/2)}{\Gamma\left(\frac{\gamma-1}{2}\right)} \left( \frac{R_{\text{eff}}}{R_E}\right)^{2-\gamma}\\
    &\times \frac{1}{b^{4(2-\gamma)}}\frac{A(\gamma, \beta; R_A, R_{\text{eff}})}{B(R_A, R_{\text{eff}})},
\end{split}
\end{align}
}
where

{\tiny
\begin{align}
\begin{split}
    A(\gamma, \beta; R_A, R_{\text{eff}}) &= \int_0^{\frac{R_A}{R_{\rm eff}}b^4} \int_{-\infty}^{\infty}dRdZ \frac{R}{(R^2+Z^2)^{\beta}}\\
    &\times \Gamma\left(4+4(2\beta-\gamma-\delta+1), ( R^2+Z^2)^{1/8} \right)\left(1- \beta \frac{R^2}{R^2+Z^2} \right),
\end{split}
\end{align}
}
and
{\tiny
\begin{eqnarray}
    B(R_A, R_{\text{eff}}) = \int_0^{\frac{R_A}{R_{\rm eff}}b^4} \int_{-\infty}^{\infty}dRdZ \frac{R}{(R^2+Z^2)^{\delta/2}}\exp\left(-(R^2+Z^2)^{1/8}\right),
\end{eqnarray}
}
with $\Gamma(.,.)$ the upper incomplete gamma function:

\begin{eqnarray}
    \Gamma(s,x) = \int_x^\infty t^{s-1}e^{-t}dt.
\end{eqnarray}

$A$ and $B$ are numerically expensive to compute. As shown in Section \ref{seq: data sample} Eq. \eqref{eq: normalisation}, $A$ and $B$ do not depend on $R_A$ and $R_{\text{eff}}$, as $B$ is constant and $A$ can be obtained  through interpolation in the $(\gamma, \beta)$-plane. We use a Gaussian Process with a Matern 5/2 kernel to avoid the untimely call to a numerical integrator. We thus have an expression of the velocity dispersion depending on the Einstein radius of GR. We will mainly focus our interest on the De Vaucouleurs deprojected luminosity profile but will compare its results to those of the power-law model (See eq. \eqref{eq: los vd pl}).

\subsubsection{Gravitational slip and screening mechanisms}

So far, we have not introduced the gravitational slip. This can be done by making the link between the observed Einstein radius, $\theta_{E, \text{obs}}$, and the one predicted by GR, $\theta_{E, \text{GR}}$, given the lens mass distribution derived from the observed velocity dispersion. The post-Newtonian variables are applied to quantify the behaviour of gravity and deviations from GR. We express the metric on cosmological scales as \cite{maCosmologicalPerturbationTheory1995},

\begin{eqnarray}\label{eq: metric}
    ds^2 = a^2(\eta)[-(1+2\Phi)d\eta^2 + (1-2\Psi) d \Vec{x}^2],
\end{eqnarray}
where $a^2(\eta)$ is the cosmological scale factor, $\eta$ the conformal time and $\Phi$ and $\Psi$ are the Newtonian and longitudinal gravitational potentials. In the weak-field limit, GR predicts $\Phi=\Psi$ making it possible to constrain possible departures from GR using the gravitational slip parameter $\gamma_{\rm PN} = \Phi/\Psi$. MG theories such as $f(R)$ \cite{sotiriouTheoriesGravity2010}, Brans-Dicke gravity \cite{koyamaCosmologicalTestsModified2016} or massive gravity \cite{schmidt-mayRecentDevelopmentsBimetric2016, enanderStrongLensingConstraints2013} all predict a difference bewteen the two potentials $\Phi \ne \Psi$. In many MG theories, $\gamma_{\rm PN}=1$ is expected at small and/or large scales due to screening effects and a limited range of the additional fifth force. Gravitational screening suppresses the additional gravitational degrees of freedom introduced by MG theories within a certain scale, in massive gravity theory refered to as the Vainshtein radius $r_V$. At large scales, the Compton wavelength of the massive graviton $\lambda_g$ sets the characteristic length of the Yukawa decay. Photons follow null geodesics, $ds^2=0$ and are thus affected by a potential $\Sigma \equiv \Phi + \Psi$ \eqref{eq: metric}. We can model the departure from GR phenomenologically,

\begin{eqnarray}
    \Sigma = [2+(\gamma_{\rm PN} -1)\epsilon(r; r_V, \lambda_g)]\Phi(r),
\end{eqnarray}
where $\epsilon$ is a slip profile parameterised by $r_V$ and $\lambda_g$. Note that the functional form of $\epsilon$ depends on the specific MG theory studied. $\epsilon=1$ corresponds to a scale-independent deviation from GR \cite{liuGalaxyscaleTestGeneral2022, caoTESTPARAMETRIZEDPOSTNEWTONIAN2017, schwabGalaxyScaleStrongLensingTests2010}, discussed in Section \ref{seq: constant grav slip}. In Refs. \onlinecite{jyotiCosmicTimeSlip2019, adiProbingGravitationalSlip2021, lianDirectTestsGeneral2022}, a step function corresponding to $\epsilon(r;r_V, \lambda_g) = \Theta(r-r_V)$ is employed. This description covers a large variety of models. The key feature is the computation of the deflection angle $\alpha(\theta)$ \cite{enanderStrongLensingConstraints2013}

\begin{eqnarray}
    \alpha = \frac{1}{c^2}\frac{D_{ls}}{D_s}\int_{-\infty}^{\infty} \nabla_{\perp}\Sigma dZ,
\end{eqnarray}
where $\nabla_{\perp}$ is the gradient perpendicular to the direction of the photon. We distinguish the deflection angle in GR and the additional contribution from the fifth force parameterised by $\gamma_{\rm PN}$, $r_V$ and $\lambda_g$:

{\small
\begin{align}
    \alpha_{\rm GR}(\theta) &=  \frac{2}{c^2}\frac{D_{ls}}{D_s} \int_{-\infty}^{\infty} \frac{\partial \Phi}{\partial R} dZ,\\
    \Delta \alpha(\theta) &= \frac{\gamma_{\rm PN}-1}{c^2}\frac{D_{ls}}{D_s} \int_{-\infty}^{\infty} \frac{\partial}{\partial R}(\epsilon(r; r_V, \lambda_g)\Phi(r))dZ.
\end{align}
}

The lens equation, with $\beta$ the source angular position, is given by

\begin{eqnarray}\label{eq: lens equation}
    \beta(\theta) = \theta - \alpha_{\rm GR}(\theta) - \Delta \alpha(\theta).
\end{eqnarray}

Setting $\beta=0$ draws a map from the observed Einstein radius, $\theta_{E, \text{obs}}$, and the one predicted by GR, $\theta_{E, \text{GR}}$. The slip profile considered in this work is

\begin{eqnarray}
    \epsilon(r; r_V, \lambda_g) = \frac{r^f}{r_V^f + r^f}e^{-r/\lambda_g},
\end{eqnarray}
where $f$ is an additional parameter tuning the sharpness of the cutoff at small scales. It models a polynomial screening at small scales and an exponential decay at large scales in keeping with bimetric gravity \cite{enanderStrongLensingConstraints2013, babichevRestoringGeneralRelativity2013}. For consistency with the latter theory, we fix $f=3$ throughout this work. In more general terms, we model $\gamma_{\rm PN}$ as radius dependent. $\epsilon$ encodes the radial profile of $\gamma_{\rm PN}$ which goes to $1$ for $r \ll r_V$ and $r \gg \lambda_g$ and the remaining degree of freedom is the maximum deviation of the gravitational slip from unity. The deflection angle associated with this screening function is

{\tiny
\begin{align}
    &\alpha_{\rm GR}(\theta) = \theta_{E, \text{GR}}^{\gamma-1}\theta^{2-\gamma},\\
    \begin{split}
    &\Delta \alpha(\theta) = \frac{\gamma_{\rm PN}-1}{2\sqrt{\pi}} \frac{\Gamma\left( \frac{\gamma}{2}\right)}{\Gamma\left( \frac{\gamma-1}{2} \right)} (\theta_{E, \text{GR}} D_L)^{\gamma-1}\theta\\
    &\times \underbrace{\int_{-\infty}^{\infty}dZ e^{-r/\lambda_g}r^{f-\gamma}\left(\left[ 1 - \frac{r}{\lambda_g (2-\gamma)}\right] \frac{1}{r_V^f+r^f} + \frac{f}{2-\gamma}\frac{r_V^f}{(r_V^f+r^f)^2}\right)}_{I(D_l \theta; \lambda_g, r_V, \gamma, f)}.
\end{split}
\end{align}
}

giving, using \eqref{eq: lens equation}:

{\tiny
\begin{align}\label{eq: generalized theta GR}
    \theta_{E, \text{GR}} = \left(\theta_{E, \text{obs}}^{1-\gamma} + \frac{\gamma_{\rm PN}-1}{2\sqrt{\pi}} \frac{\Gamma(\gamma/2)}{\Gamma((\gamma-1)/2)} D_l^{\gamma-1}  I(D_l \theta_{E, \text{obs}}; \lambda_g, r_V, f, \gamma)\right)^{\frac{1}{1-\gamma}}.
\end{align}
}

\subsection{Data sample}\label{seq: data sample}

In this work, we recovered a subsample of the data used in Ref. \onlinecite{chenAssessingEffectLens2019} (See Table A1 in the latter), having the benefit of being a recently compiled dataset for strong lensing. It contains 130 galaxy-scale SGL systems selected to approximately comply with the assumption of spherical symmetry via the following criteria:

\begin{itemize}
    \item The lens galaxy should be an Early-Type Galaxy with E/S0 morphologies.
    \item The lens galaxy should not have significant substructure or close massive companion.
\end{itemize}

Among those 130 systems, 57 comes from the SLACS survey \cite{boltonSloanLensACS2008}, 38 from the SLACS extension SLACS for the Masses \cite{shuSloanLensACS2017}, 21 from the BELLS \cite{brownsteinBOSSEmissionLineLens2012} and 14 from the BELLS for GALaxt-Ly$\alpha$ EmitteR sYstems (BELLS GALLERY, \cite{shuBOSSEmissionlineLens2016}).\\

This dataset provides the following information of relevance to compute the theoretical velocity dispersion from equations \eqref{eq: los vd pl} or \eqref{eq: los vd dv}:

\begin{itemize}
    \item $z_l$, the lens redshift.
    \item $z_s$, the source redshift.
    \item $\theta_{E, \text{obs}}$, the observed Einstein angle.
    \item $\sigma_{\text{ap}}$, the velocity dispersion of the lens galaxy in the corresponding spectroscopic aperture.
    \item $\Delta \sigma_{\text{ap}}$, the associated measurement error.
    \item $\theta_{\text{ap}}$, the spectroscopic aperture angular radius.
    \item $\theta_{\text{eff}}$, the half-light angular radius of the lens galaxy.
    \item $\delta$, power-law index of the luminosity density\footnote{When required this index has been fitted on the high-resolution HST imaging data for the galaxies in our sample, see \cite{chenAssessingEffectLens2019} for details.}.
\end{itemize}

To take into account the effect of the aperture size on the measurements of the velocity dispersions $\sigma_{\text{ap}}$, we normalise all velocity dispersion to the typical physical aperture $\theta_{\text{eff}}/2$:

\begin{align}\label{eq: normalisation}
    \sigma_{\parallel}^{\text{obs}} = \sigma_{\text{ap}}\left(\frac{\theta_{\text{eff}}}{2\theta_{\text{ap}}} \right)^{\eta}.
\end{align}

We adopt the best-fit value of $\eta = -0.066 \pm 0.035$ from Ref. \cite{cappellariSAURONProjectIV2006}. The total uncertainty of $\sigma_{\parallel}^{\text{obs}}$ can thus be written \cite{liuGalaxyscaleTestGeneral2022}:

\begin{align}\label{eq: uncertainty}
    (\Delta \sigma_{\parallel}^{\text{tot}})^2 = \left[\frac{\Delta \sigma_{\text{ap}}^2}{\sigma_{\text{ap}}^2}+\Delta \sigma_{\text{sys}}^2+ \left[\ln \left( \frac{\theta_{\text{eff}}}{2\theta_{\text{ap}}}\right) \Delta \eta \right]^2 \right] (\sigma_{\parallel}^{\text{obs}})^2,
\end{align}

where we include a systematic error of $\Delta \sigma_{\text{sys}}$, e.g., taking into account possible extra mass contribution
from matter along the LOS \cite{jiangBaryonFractionsMasstoLight2007}. Previous work introduced a systematic error of $3\%$. To assess the uncertainty linked to the mass model, we run an MCMC analysis with $\gamma_{\rm PN}=1$ and $\Delta \sigma_{\rm sys}$ as an additional parameter. The fitted value for the systematic error is $\Delta \sigma_{\rm sys} = 9.52 \pm 0.01\%$ larger than the one used in previous work. In what follows, the latter value for the systematic error is used. The corresponding theoretical prediction of the velocity dispersion is obtained by evaluating equations \eqref{eq: los vd pl} and \eqref{eq: los vd dv} at $R_A = R_{\text{eff}}/2$,

\begin{eqnarray}
    \sigma_{\parallel}^{\text{th}} = \sigma_{\parallel}(\leq R_{\text{eff}}/2).
\end{eqnarray}

In our analysis, we assume a Gaussian likelihood:

\begin{eqnarray}\label{eq: likelihood}
    \mathcal{L} \propto e^{-\chi^2/2},
\end{eqnarray}

where

\begin{eqnarray}
    \chi^2 = \underset{i=1}{\overset{N}{\sum}} \left( \frac{\sigma_{\parallel, i}^{\text{th}}-\sigma_{\parallel, i}^{\text{obs}}}{\Delta \sigma_{\parallel, i}^{\text{tot}}}\right)^2,
\end{eqnarray}

with $N$ being the number of SGL systems. In the following analysis, we derive the posterior probability distributions of model parameters using an affine-invariant Markov Chain Monte Carlo (MCMC) Ensemble sampler (emcee; \cite{foreman-mackeyEmceeMCMCHammer2013}).

\subsection{Cosmological model}

In equations \eqref{eq: los vd pl} and \eqref{eq: los vd dv}, we use a $\Lambda$CDM cosmology such that the angular distance between redshift $z_1$ and $z_2$ is given by

\begin{eqnarray}
    D(z_1, z_2; H_0, \Omega_m) = \frac{c}{H_0(1+z_2)}\int_{z_1}^{z_2}\frac{dz}{E(z;\Omega_m)},
\end{eqnarray}

\begin{eqnarray}
    E(z;\Omega_m) = \sqrt{\Omega_m (1+z)^3 + (1-\Omega_m)},
\end{eqnarray}

where $H_0=67.37$ km/s/Mpc and $\Omega_m=0.315$ \cite{planckcollaborationPlanck2018Results2020}.\\

It is common in the literature to use a cosmology-independent approach to compute the angular distance, usually using Type Ia supernovae data to get luminosity distances up to redshift $z \simeq 2$ (See Refs. \cite{lianDirectTestsGeneral2022, chenAssessingEffectLens2019, liuTestingCosmicCurvature2020}). We chose not to adopt such an approach and argue that the cosmological model has only negligible influence on the results. Also, since we only need ratios of angular distances $D_s/D_{ls}$, the results do not depend on the Hubble constant $H_0$. As evident from Figure 2 in \cite{chenAssessingEffectLens2019}, the influence of $\Omega_m$ on the ratio is quite small, at least for lenses at small redshift. Finally, Figure 4 and Table 1 in \cite{liuGalaxyscaleTestGeneral2022} show that the use of distance calibration yields only minor modifications to the fitted values. The reader should nevertheless keep in mind that using $\Lambda$CDM to measure distances and constrain GR should be considered an approximation employed for simplicity, motivated by the fact that a polynomial fit of Type Ia supernova data will only yield small differences in the estimation of angular distances.

\subsection{Model parameters and priors}\label{seq: model and prior}

We run MCMC chains to fit the gravitational slip, $\gamma_{\rm PN}$, the mass density slope $\gamma$, and the velocity anisotropy, $\beta$. The gravitational slip is our main interest but it requires accurate constrains on the lens mass model \cite{caoLimitsPowerlawMass2016, caoTESTPARAMETRIZEDPOSTNEWTONIAN2017}. $\gamma$ corresponds to a common total density slope across our sample. We adopt flat priors for $\gamma_{\rm PN}$ and $\gamma$ on sufficiently wide ranges. We cannot independently measure $\beta$ for individual lensing system with the spectroscopic data available. The latter is thus considered as a nuisance parameter, and therefore needs an informative prior.\\

\subsubsection{Prior on the velocity anisotropy}

A truncated Gaussian prior on the velocity anisotropy $\beta$ is commonly used with $\beta = 0.18 \pm 0.13$ truncated at $[\overline{\beta} - 2\sigma_{\beta}, \overline{\beta} + 2\sigma_{\beta}]$ \cite{chenAssessingEffectLens2019, schwabGalaxyScaleStrongLensingTests2010, lianDirectTestsGeneral2022, liuGalaxyscaleTestGeneral2022}. This constraint is obtained from a well-studied sample of nearby elliptical galaxies \cite{gerhardDynamicalFamilyProperties2001}. We assess the influence of the prior on $\beta$ by introducing a new prior based on the most recent dynamical data of E/S0 galaxies from the combined analysis of the Dynamical and stellar Population (DynPop) for the MaNGA survey in the final SDSS Data Release 17 \cite{zhuMaNGADynPopQualityassessed2023}. It contains dynamical data of $\sim 10^4$ galaxies in the local universe analysed using the axisymmetric Jean Anisotropic Modelling (JAM) method. The latter is based on the Jeans equation with the velocity anisotropy $\beta$ as a parameter. In line with our spherically symmetric assumption, we consider the models using $\text{JAM}_{\text{sph}}$. We moreover only use the NFW and gNFW mass models since the mass-follows-light and the fixed NFW do not recover the density profiles very well. Finally, to avoid bias, we only select E/S0 galaxies using the method in Ref. \onlinecite{zhuMaNGADynPopIII2023}. To only select the most reliable data, we further impose

\begin{eqnarray}
    |\beta_{NFW} - \beta_{gNFW}| < 0.05.
\end{eqnarray}

\begin{figure}
\includegraphics[width = 1\linewidth]{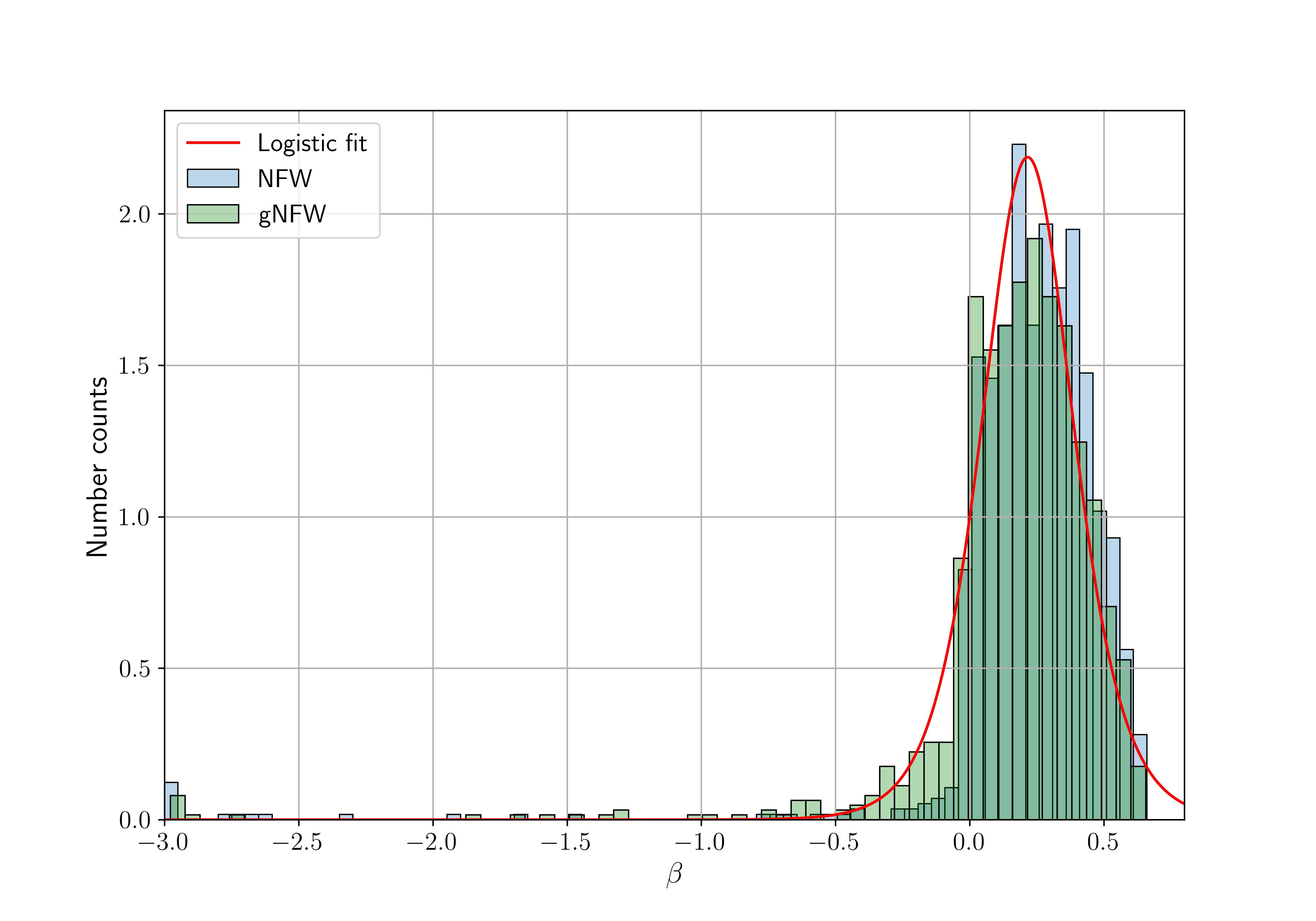}
\caption{\label{fig: beta_prior_image} Distribution of the anisotropy parameter $\beta$ from MaNGA DynPop modelling \cite{zhuMaNGADynPopQualityassessed2023}. The blue and green histograms correspond to the distribution obtained with an NFW and a gNFW model, respectively. The red solid curve corresponds to the best-fit of the histograms obtained with a logistic distribution (See eq. \eqref{eq: logistic}).}
\end{figure}

The threshold has been chosen to ensure a reasonable trade-off between the amount of data and the quality of the fit of $\beta$. Our final sample contains 1136 galaxies to which we fit several distributions in order to find the most realistic prior. We finally chose a logistic prior to be compared with the histograms of our data in Figure \ref{fig: beta_prior_image},

\begin{align}\label{eq: logistic}
    f(x;\mu, s) = \frac{e^{-(x-\mu)/s}}{s(1+e^{-(x-\mu)/s^2})},
\end{align}

where $f$ is the logistic's density and $\mu$ and $s$ are the location and scale parameters fitted to the histograms. The logistic's wings are wider than the Gaussian's so it will allow the MCMC analysis to allow for larger range of values for the velocity anisotropy.  The logistic is truncated at $3\sigma$ to prevent $\beta$ from taking nonphysical values, e.g., $\beta>1$.\\

\subsubsection{Grid analysis of screening mechanisms}

In Section \ref{seq: grav slip screening}, we will introduce screening mechanisms by performing the fit for various values of the Compton wavelength of the theory, $\lambda_g$. The latter will span order of magnitudes from the pc scale to the Gpc scale. Motivated by bimetric theory, we make the Vainshtein radius $r_V$ dependent of $\lambda_g$ and the mass of the lens galaxy \cite{enanderStrongLensingConstraints2013},

\begin{eqnarray}\label{eq: vainshtein}
    r_V = (r_S \lambda_g^2)^{1/3},
\end{eqnarray}

where $r_S$ is the Schwarzschild radius of the lens considered given by the mass inside its Einstein radius $\theta_{E, \text{obs}}$. The Vainshtein radius is therefore different for each galaxy in our sample. Varying $\lambda_g$ explore regimes where the lenses in our samples are screened or unscreened explaining why we rather perform sampling of the gravitational slip for various $\lambda_g$ rather than including it in our parameters. The former case allows to study the dependency of the constraints of $\gamma_{\rm PN}$ on $\lambda_g$ whereas the latter would not sample the full range of $\lambda_g$.

\section{Results and Discussion}\label{seq: results and discussion}

We first assess the influence of the lens mass model in the case of a scale independent gravitational slip in Section \ref{seq: constant grav slip}. We then study the constraints on a scale dependent gravitational slip (Section \ref{seq: grav slip screening}) and discuss the results in Section \ref{seq: discussion}.

\begin{figure*}
    \includegraphics[width = 0.45\linewidth]{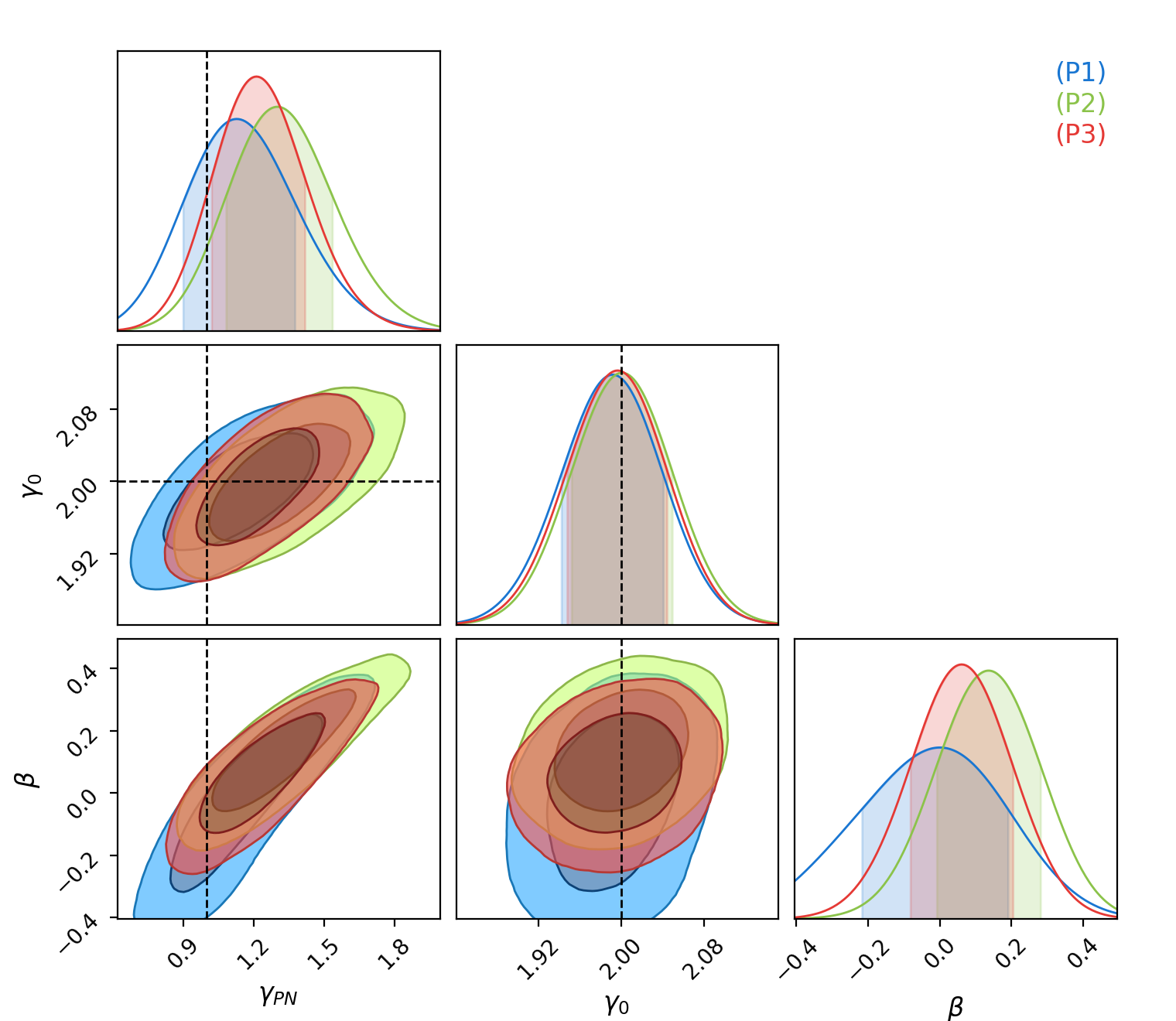}
    \includegraphics[width = 0.45\linewidth]{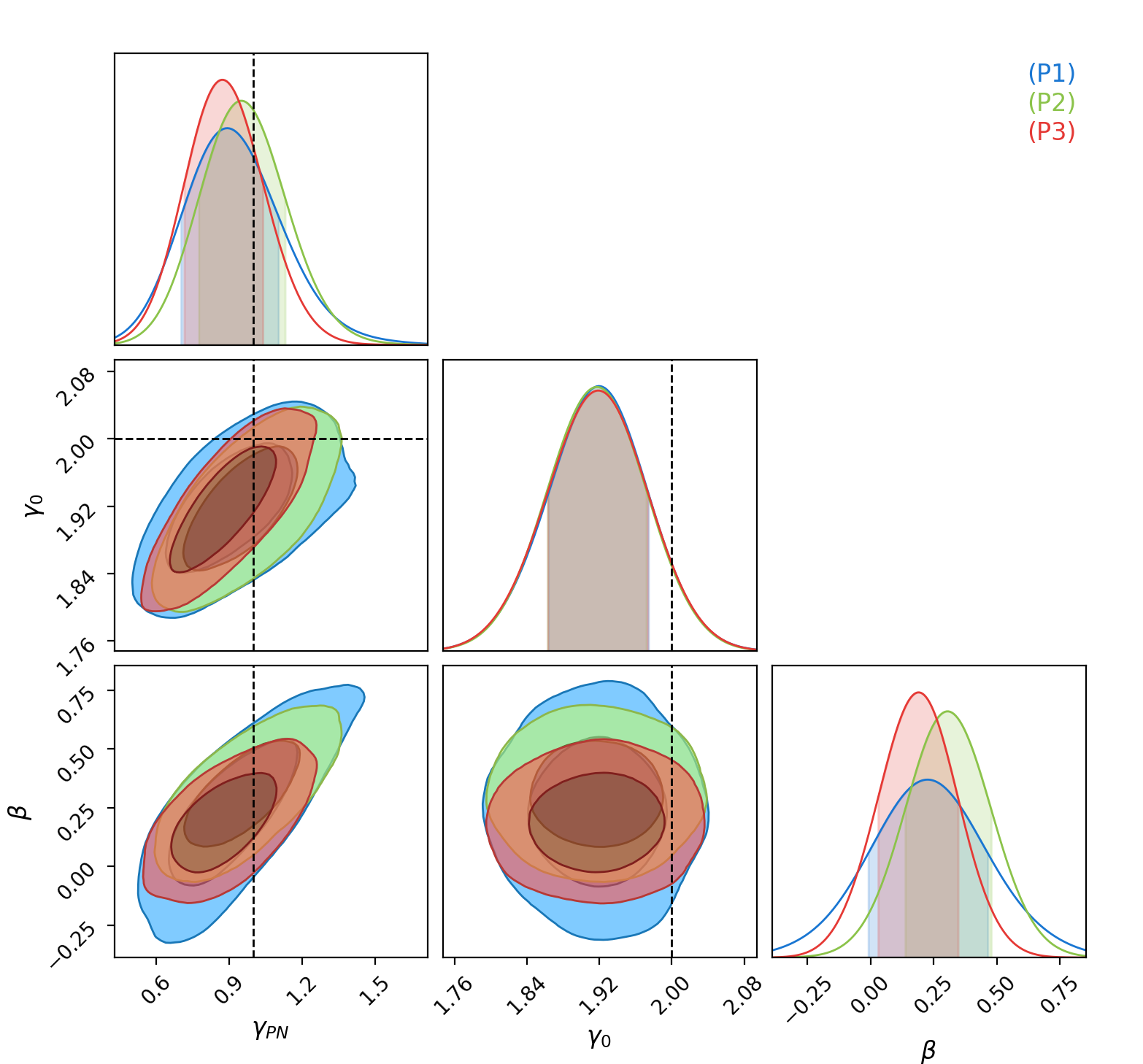}
    \caption{1D and 2D marginalised probability distribution at the $1\sigma$ and $2\sigma$ confidence level for the gravitational slip parameter $\gamma_{\rm PN}$ and the lens mass model parameters in the case of power-law profile (\textit{left panel}) or a De Vaucouleurs profile (\textit{right panel}) for the luminosity density. The dashed lines represent $\gamma_{\rm PN}=1$ predicted by GR and $\gamma=2$ expected for a Singular Isothermal Sphere.}
    \label{fig: M1_pl}
\end{figure*}

\begin{table*}
\caption{\label{tab: prior_results}The 1D marginalized limit (68\% confidence regions) for model parameters constrained from the truncated sample with 130 SGL systems with various priors on $\beta$ for two different models of the luminosity density. The bottom panel corresponds to the case of GR where we fit the lens mass model to the data with $\gamma_{\rm PN}=1$.}
\begin{ruledtabular}
\begin{tabular}{cccccccc}
 Luminosity density & Prior on $\beta$ & Parameters & & & $\chi^2_{\text{min}}$ & AIC & BIC\\ \hline
 Power-law& (P1) &$\gamma_{\rm PN} = 1.14^{+0.22}_{-0.18}$& $\gamma = 1.99^{+0.04}_{-0.04}$ & $ \beta = -0.02^{+0.16}_{-0.19}$& 156.2 & 162.2 & 170.8  \\
 Power-law& (P2) & $\gamma_{\rm PN} = 1.31^{+0.20}_{-0.17}$& $\gamma = 2.00^{+0.04}_{-0.04}$ & $ \beta = 0.14^{+0.12}_{-0.12}$ & 158.4 & 164.4 & 173.0\\
 Power-law& (P3) & $\gamma_{\rm PN} = 1.22^{+0.17}_{-0.15}$& $\gamma = 2.00^{+0.04}_{-0.04}$ & $ \beta = 0.06^{+0.12}_{-0.12}$ & 157.2 & 163.2 & 171.8\\
 De Vaucouleurs& (P1) & $\gamma_{\rm PN} = 0.90^{+0.18}_{-0.14}$& $\gamma = 1.92^{+0.04}_{-0.05}$ & $ \beta = 0.23^{+0.19}_{-0.19}$& 146.8 & 152.8 & 161.4\\
 De Vaucouleurs& (P2) & $\gamma_{\rm PN} = 0.96^{+0.15}_{-0.14}$& $\gamma = 1.92^{+0.04}_{-0.05}$ & $ \beta = 0.31^{+0.14}_{-0.13}$ & 146.8 & 152.8 & 161.4\\
 De Vaucouleurs & (P3) & $\gamma_{\rm PN} = 0.88^{+0.14}_{-0.13}$& $\gamma = 1.92^{+0.05}_{-0.05}$ & $ \beta = 0.19^{+0.13}_{-0.13}$ &146.8 & 152.8 & 161.4\\
 \hline
 Power-law& (P1) &$\gamma_{\rm PN} = 1$ & $\gamma = 1.97^{+0.03}_{-0.03}$ & $ \beta = -0.15^{+0.11}_{-0.08}$& 154.8 & 158.8 & 169.4  \\
 De Vaucouleurs & (P1) & $\gamma_{\rm PN} = 1$& $\gamma = 1.94^{+0.04}_{-0.03}$ & $ \beta = 0.30^{+0.14}_{-0.11}$& 147.1 & 151.1 & 161.7\\
\end{tabular}
\end{ruledtabular}
\end{table*}

\subsection{Constant gravitational slip}\label{seq: constant grav slip}

For $\epsilon(r; r_V, \lambda_g)=1$, the relation between $\theta_{E, \text{GR}}$ and $\theta_{E, \text{obs}}$ is obtained from \eqref{eq: generalized theta GR} with $r_V=0$ and $\lambda_g \rightarrow \infty$,

\begin{eqnarray}
    \theta_{E, \text{GR}} = \theta_{E, \text{obs}}\left( \frac{\gamma_{\rm PN}+1}{2}\right)^{-\frac{1}{\gamma-1}}.
\end{eqnarray}

We perform the analysis for a power-law luminosity density and a De Vaucouleurs luminosity density (Figure \ref{fig: M1_pl}). We moreover study the influence of the prior on the velocity anisotropy $\beta$ by considering three priors:

\begin{itemize}
    \item[(P1)]Logistic distribution fitted to MaNGA DynPop dynamical data (See Section \ref{seq: model and prior} and Figure \ref{fig: beta_prior_image}) truncated at $[\mu -3\sigma, \mu + 3\sigma]$ with $(\mu, \sigma)=(0.22, 0.2)$.
    \item[(P2)]Truncated Gaussian with $(\mu, \sigma)=(0.3, 0.14)$ between $[\mu -3\sigma, \mu + 3\sigma]$.
    \item[(P3)]Truncated Gaussian with $(\mu, \sigma)=(0.18, 0.13)$ between $[\mu -2\sigma, \mu + 2\sigma]$ used in previous work.
\end{itemize}

The results are summarised in Table \ref{tab: prior_results}. We use the Akaike Information Criterion (AIC) \cite{akaikeNewLookStatistical1974} and the Bayesian Information Criterion (BIC) \cite{schwarzEstimatingDimensionModel1978} as statistical criterion for model selection

\begin{eqnarray}
    \text{AIC} &=& 2k+\chi^2_{\text{min}},\\
    \text{BIC} &=& k\ln(N) + \chi^2_{\text{min}},
\end{eqnarray}

where $k$ is the number of parameters and $N$ the number of data points. They award models with few parameters giving good fits to the data. Here, models containing additional parameters for either screening or the lens mass are penalized in terms of the IC's, unless they supply significant better fits compared to the baseline model. Only the relative difference in AIC and BIC is relevant to favor a model over another.\\

The best-fit values of a constant $\gamma_{\rm PN}$ are all consistent with GR at the 68\% confidence level. Particularly, in the case of a logistic prior (P1), we find a best-fit value of the gravitational slip of $\gamma_{\rm PN} = 1.14^{+0.22}_{-0.18}$ in the case of power-law luminosity densities and $\gamma_{\rm PN} = 0.90^{+0.18}_{-0.14}$ in the case of deprojected De Vaucouleurs luminosity densities.\\

The gravitational slip and the velocity anisotropy are positively correlated and the prior on $\beta$ can influence the fitted value of $\gamma_{\rm PN}$ in the case of a power-law luminosity density (See Figure \ref{fig: M1_pl}). Our choice of prior based on recent dynamical data \cite{zhuMaNGADynPopQualityassessed2023} is slightly favored upon commonly used Gaussian priors but the IC's is not significantly better. We however underline that the posterior of the velocity anisotropy $\beta$ is biased towards low values in the case of a logistic prior. The fitted value of the gravitational slip $\gamma_{\rm PN}$ is therefore prior dependent. The best-fit values of the gravitational slip in the case of a deprojected De Vaucouleurs profile depend less on the prior choice for $\beta$. We find $\gamma_{\rm PN}=0.90$,  $0.96$ and $0.88$ for priors (P1), (P2) and (P3), results agreeing at the 68\% confidence level and being consistent with GR. We further note that the De Vaucouleurs luminosity profile improves the AIC with a value of $146.8$ against $156.2$ in the power-law case using a logistic prior on $\beta$. Hereafter, we use the logistic prior on the velocity anisotropy $\beta$ since it represents well the most recent dynamical data. In the GR case ($\gamma_{\rm PN}=1$) with this logistic prior, the fitted lens mass model gives an $\text{AIC}_{\rm GR, DV}=151.1$ and $\chi^2_{\rm GR, DV}=147.1$ for a De Vaucouleurs luminosity profile. The GR case is favored over the constant gravitational slip case, since adding a constant gravitational slip does not give a significantly better representation of the data. The GR-case will serve as our reference model. In the case of a power-law luminosity density, we get $\text{AIC}_{\rm GR, PL}=158.8$ and $\chi^2_{\rm GR, PL}=154.8$ which performs better than the case with a gravitational slip parameter. \\

We underline that the value of $\gamma$ is positively correlated with the gravitational slip. Our result $\gamma\in[1.9, 2.1]$ is consistent with previous studies fitting values of E/S0 galaxies density slope close to the Singular Isothermal Sphere (SIS) value of $\gamma=2$ \cite{liStronglensingMeasurementTotalmassdensity2018}.

\begin{figure}
    \includegraphics[width = 1\linewidth]{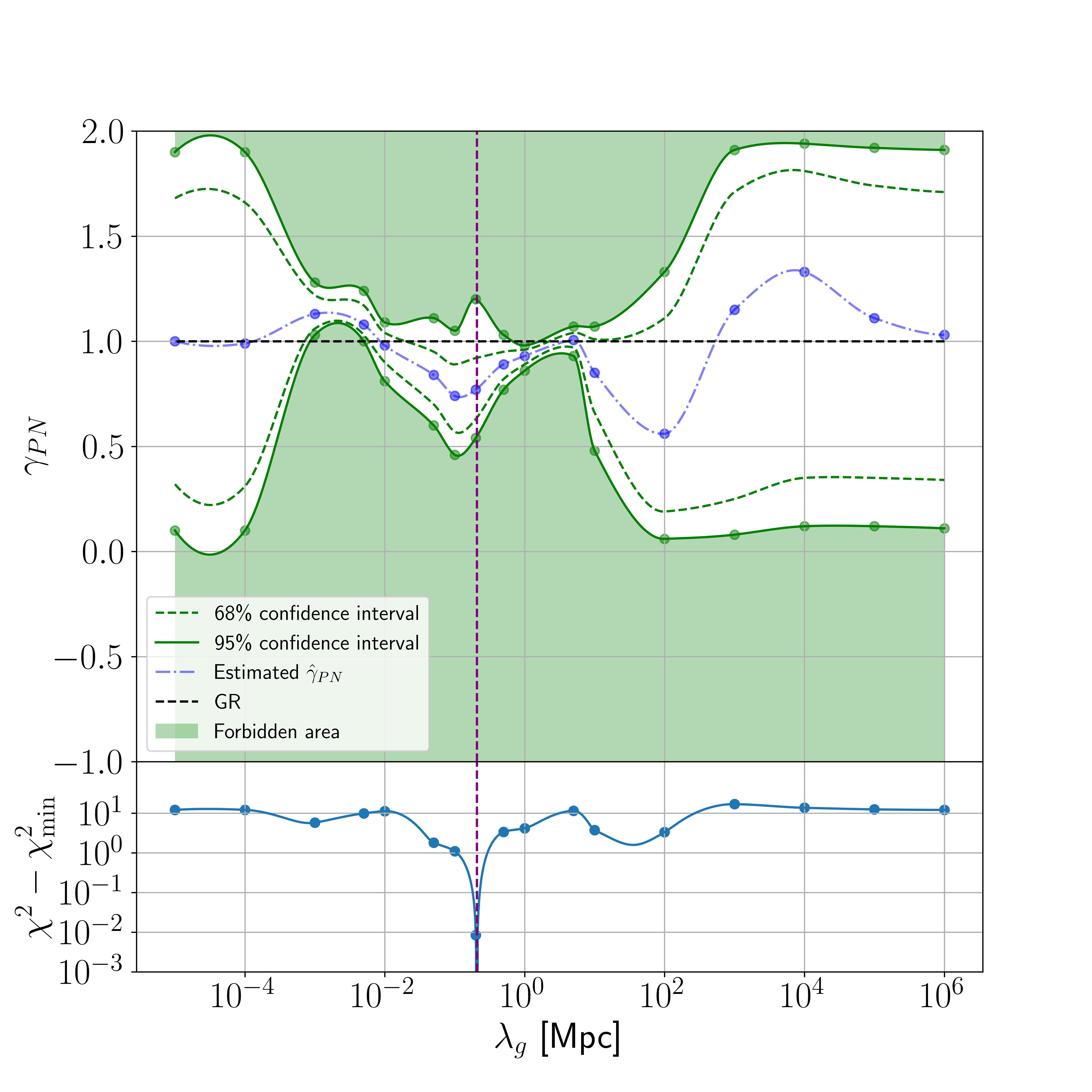}
    \caption{Fitted values of $\gamma_{\rm PN}$ for various Compton wavelength $\lambda_g$ using a De Vaucouleurs luminosity profile. The upper panel shows the evolution of the estimated $\gamma_{\rm PN}$ as well as its confidence interval at the 68\% and 95\% levels. Shaded areas correspond to regions of phase space ruled out by our constraints at the 95\% confidence level. The lower panel shows the corresponding value of the $\chi^2-\chi^2_{\text{min}}$ for each Compton wavelength. The dashed purple line corresponds to the minimum of the $\chi^2_{\rm min}=134.9$.}
    \label{fig: compton_M1_pl}
\end{figure}

\subsection{Gravitational slip under screening}\label{seq: grav slip screening}

We now introduce a scale dependent slip parameterised by the Compton wavelength $\lambda_g$. The Vainshtein radius is computed using equation \eqref{eq: vainshtein}. We fit the gravitational slip and the lens mass parameters for values of the Compton wavelength spanning from pc to Gpc scales. Our interest here is how constraints on $\gamma_{\rm PN}$ evolve with the Compton wavelength $\lambda_g$. Figure \ref{fig: compton_M1_pl} shows the 95\% confidence region of $\gamma_{\rm PN}$ depending on $\lambda_g$ for a deprojected De Vaucouleurs luminosity density only. As we can see in the bottom panel of figure \ref{fig: compton_M1_pl}, there are two competing local $\chi^2$-minima for $\lambda_g \sim 0.2$ Mpc and $\lambda_g \sim 100$ Mpc. Note that the contour plot obtained with the Compton wavelength $\lambda_g$ as a free parameters would look different since the two local minima correspond to slightly different best-fit values for the gravitational slip and the samples are drawn from different region of phase space. Gridding over $\lambda_g$'s allows for an analysis of the degeneracy between the Compton wavelength and the gravitational slip. \\

The dependence of the gravitational slip on the Compton wavelength allow us to draw qualitative conclusions. We first highlight the inability of our model to constrain $\gamma_{\rm PN}$ for $\lambda_g \leq 10^{-4}$ Mpc and $\lambda_g \geq 10^3$ Mpc. In the latter case, the Vainshtein radius for a galaxy of mass $M \sim 10^{11} M_{\odot}$ is of the order $r_V \sim 10^3-10^4$ kpc. As a result, lens galaxies in our sample are completely screened from fifth force lensing effects. Analogously, for Compton wavelengths below $\sim 100$ parsec, Einstein radii $\sim 10$ kpc correspond to large numbers of e-folds of the fifth force Yukawa decay. In both regimes, we end up fitting models effectively equivalent to the reference GR case. We note some discrepancies from GR when fixing the Compton wavelength to order $\lambda_g \sim 10^{-2}$ Mpc and $\lambda_g \sim 1$ Mpc. However, for these $\lambda_g$ and values between, the obtained constraints on the gravitational slip have no statistically significant departures from GR. 


Quantitatively, for intermediate Compton wavelength $\lambda_g$, various constraints on the gravitational slip are obtained but computing the $\chi^2$ hints at the most likely configuration. The best fit is obtained for $\lambda_g \sim 0.2$ Mpc i.e $r_V \sim 1$ kpc. The corresponding gravitational slip $\gamma_{\rm PN} = 0.77^{+0.43}_{-0.23}$ at the 95\% confidence level with $\chi^2_{\text{min}} = 134.9$ yielding $\text{AIC}_{\text{min}}=142.9$. Including screening mechanisms provides a better fit to the data but the result is consistent with GR at the 95\% confidence level. Note that the bottom panels of Figure \ref{fig: compton_M1_pl} shows that $\lambda_g \sim 100$ Mpc presents a local minimum with a $\chi^2 = 136.5$, slightly larger than for $\lambda_g \sim 0.2$ Mpc (See Table \ref{tab: results}). 
It appears that the $\text{AIC}$ is significantly decreased when we take screening effects into account. Screening mechanisms modify the shape of the likelihood used in the GR case adding sharp variations of the $\chi^2$ sensitive to both $\gamma_{\rm PN}$ and the lens mass model $(\gamma, \beta)$. However, the likelihood only slightly varies in some direction in the $(\gamma_{\rm PN}, \gamma)$-plane up to the GR case where the $\chi^2_{\text{GR}}$ amounts to $\sim$147 explaining the size of the error bars on the gravitational slip. This phenomenon will be further discussed in section \ref{seq: discussion}.

We finally underline that the tightening of the constraints for $\lambda_g \sim 1$ Mpc and $\lambda_g \sim 10^{-3}$ Mpc correspond to the cases where the Vainshtein radius and the Compton wavelength cross the typical Einstein radii in our samples, respectively. As a result, only part of the systems are screened yielding tighter constraints on the gravitational slip.

\begin{table}
\caption{\label{tab: results}The 1D marginalized limit of the gravitational slip constrained from the truncated sample with 130 SGL systems with confidence regions at the 68\% confidence level for relevant Compton wavelengths $\lambda_g$. The $\Delta\text{AIC}$ is computed between the best-fit values reported and the $\text{AIC}$ obtained in GR $\chi^2_{\rm GR}=147.1$ and $\text{AIC}_{\text{GR}} = 151.1$.}
\begin{ruledtabular}
\begin{tabular}{cccc}
$\lambda_g$ [Mpc] & Grav. slip $\gamma_{\rm PN}$ & $\chi^2_{\text{min}}$ & $\Delta \text{AIC}_{\text{GR}}$\\ \hline
 0.2 & $0.77^{+0.25}_{-0.14} $ & 134.9 & 8.2\\
 100 & $0.56^{+0.45}_{-0.35} $ & 136.5 & 6.6\\
\end{tabular}
\end{ruledtabular}
\end{table}

\subsection{Discussion}\label{seq: discussion}

\begin{figure}
    \includegraphics[width = 0.95\linewidth]{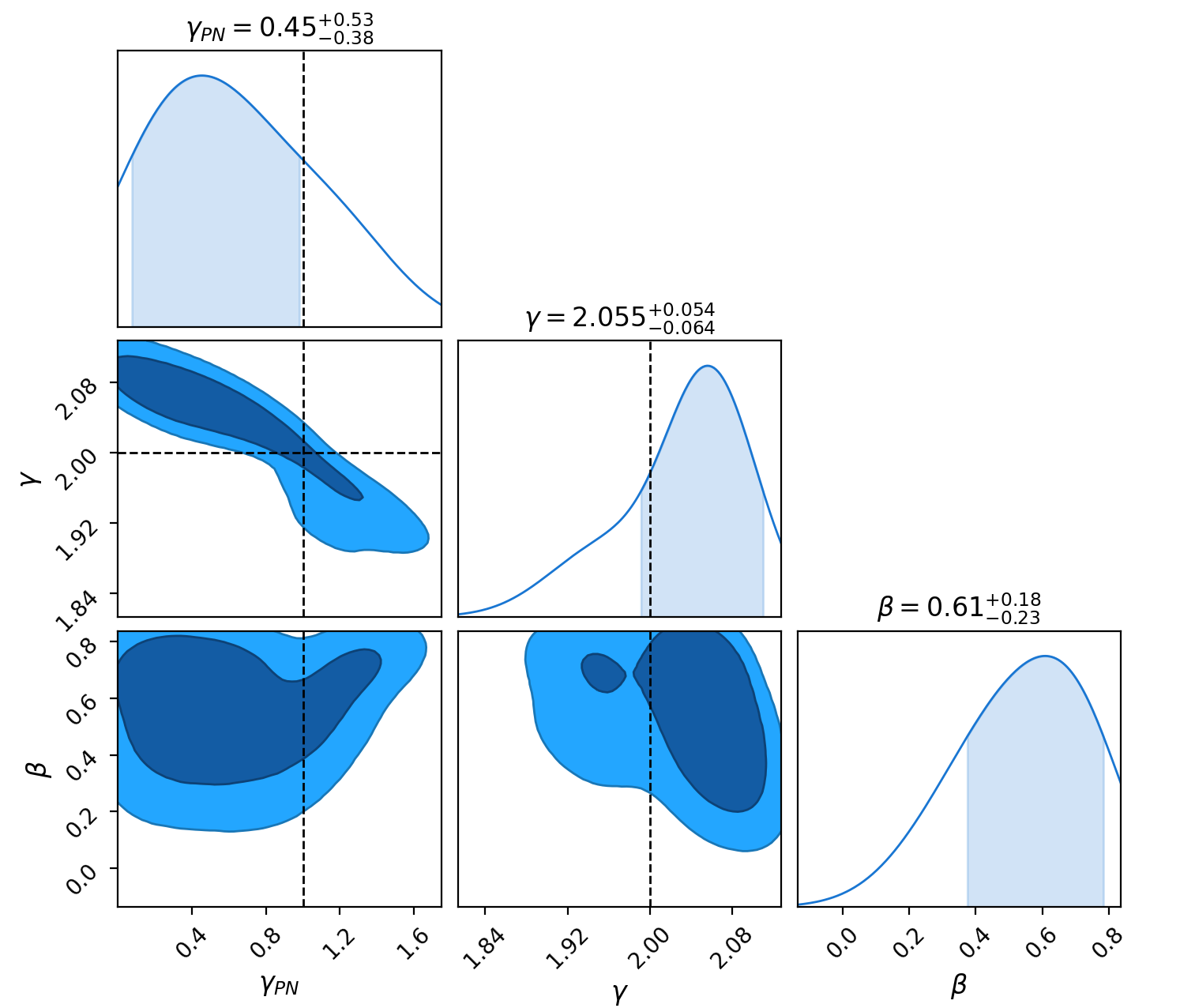}
    \caption{Confidence regions at the 68\% and 95\% confidence level of the fitted parameters for a Compton wavelength $\lambda_g = 100$ Mpc using a De Vaucouleurs luminosity density.}
    \label{fig: contour_100Mpc}
\end{figure}

\begin{figure}
    \includegraphics[width = 0.95\linewidth]{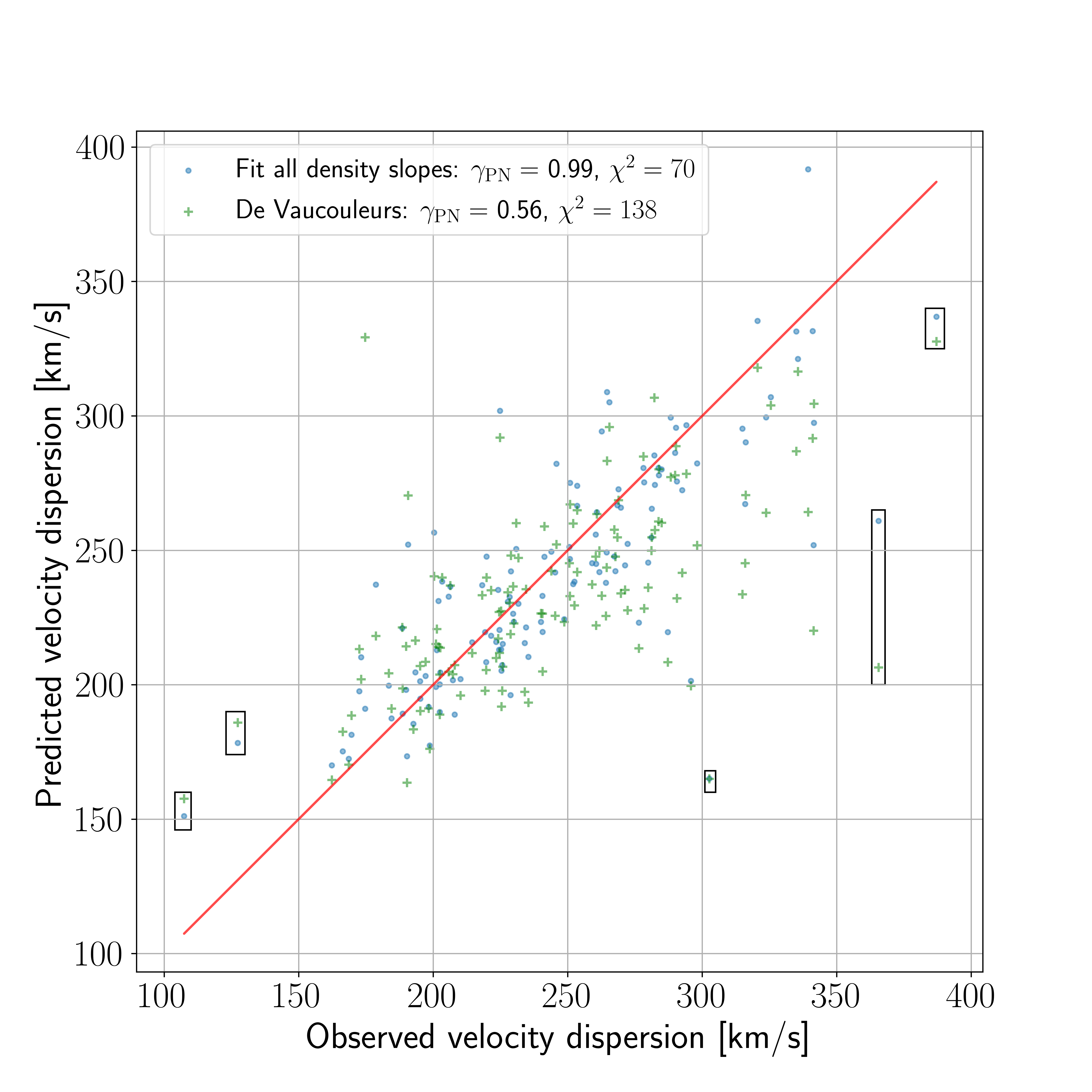}
    \caption{Scatter plot of the predicted velocity dispersion by the model against the observed velocity dispersion for $\lambda_g = 100 \text{ Mpc}$. The red solid line correspond to the ideal case where the model fits perfectly the observations. The blue dots correspond to a model where we fit a power-law index $\gamma$ for each lens system in our sample. Green crosses are obtained with a single total density slope $\gamma$. Black boxes correspond to empirically identified outliers listed in Table \ref{tab: outlier galaxies}. They were selected as manifest outliers in both the De Vaucouleurs model and the model where all density slopes are fitted.}
    \label{fig: compare_fit}
\end{figure}

Our above results present no statistically significant departure from GR, except for possible hints at $\lambda_g \sim 1$ kpc and $\lambda_g \sim 1$ Mpc. However, these Compton wavelengths are not favored in terms of the quality of their fits, or $\chi^2_{\rm min}$, meaning that if the Compton wavelength was fitted as a parameter in the MCMC analysis, the obtained contours would not include those values of the Compton wavelength. 

The obtained results however present important lessons regarding the importance of systematic uncertainties. These systematic uncertainties could be linked to the dependency of the gravitational slip on the lens mass model. In this work, we fit a common total-density slope $\gamma$ for all the lens galaxies, possibly a too simplified approximation. Figure \ref{fig: contour_100Mpc} shows confidence regions for a Compton wavelength $\lambda_g = 100$ Mpc, from which it is evident that the gravitational slip $\gamma_{\rm PN}$ is negatively correlated with the total density slope $\gamma$. This degeneracy explains the width of error bars even though the likelihood have important variations along $\gamma_{\rm PN}$. This degeneracy could be broken by having independent constraints on the total-density slope from more detailed lens mass modelling, and shows that the lens mass model is a key feature to obtain good constraints on the gravitational slip.\\ 

To further assess the influence of the lens mass model on the best-fit value of the gravitational slip, we run an MCMC analysis where the total density slope of each galaxy in our sample is a model parameter for a Compton wavelength $\lambda_g=100$ Mpc. Together with the gravitational slip and the velocity anisotropy, we thus fit 132 parameters where we assume a De Vaucouleurs luminosity density and a logistic prior on $\beta$. By doing so, we are able to study whether fitting a common matter density slope is a good assumption. Figure \ref{fig: compare_fit} presents a scatter plot of the fitted velocity dispersion against the observed one for the model with a common $\gamma$ and the case of different $\gamma$'s for each lens. It first appears that the case where all $\gamma$'s are free performs better than the case studied in this work with a $\chi^2_{\text{min}}$ of $\sim$70 against $\sim$135 even though the IC's are worse. Moreover, it measures no departure from GR with a best-fit gravitational slip of $\gamma_{\rm PN} = 0.99^{+0.027}_{-0.033}$. The latter shows that a single total density slope $\gamma$ poorly takes into account outliers (see black boxes in Figure \ref{fig: compare_fit}) listed in Table \ref{tab: outlier galaxies}. Most of them comes from BELLS survey and Ref. \onlinecite{liStronglensingMeasurementTotalmassdensity2018} fitted the total density slope for those lenses. We note that for each of those outliers, the power-law index $\gamma$ is either poorly constrained or deviates significantly from $\gamma=2$ which is the mean power-law index fitted in Ref. \onlinecite{liStronglensingMeasurementTotalmassdensity2018}. Correctly constraining the lens mass model is therefore key to find an unbiased estimate of the gravitational slip. Previous work added extra degree of freedoms using dependency on the lens redshift or its surface density \cite{chenAssessingEffectLens2019, liuGalaxyscaleTestGeneral2022}. Those correlations are however not evident in MaNGA DynPop data \cite{zhuMaNGADynPopQualityassessed2023} and should be used with caution.\\

Fitting the power-law index $\gamma$ for each lens, convergence of the MCMC analysis is difficult to assess and, even though we were able to reduce the $\chi^2$ with this method, it is likely that the lens mass model has not converged for every lens in our sample\footnote{With so many parameters, the curse of dimensionality does not allow us to know if we sufficiently explored parameter space.}. Further investigations of the lens mass modelling should lead to significant improvement in the measurement of the gravitational slip. We suggest two directions to further investigate gravitational slip constraints. The first being an approach where we ensure a good control of the mass model. To do so, we select a small number of systems for which we have the required photometric and spectroscopic data to constrain the lens mass model individually for each system, e.g, using packages like lenstronomy \cite{birrerLenstronomyMultipurposeGravitational2018}. We argue that this approach could prevent the presence of outliers in our dataset and yields more reliable constraints on the slip by lifting the degeneracy between the gravitational slip and the total-density slope. Second, it could be worthwhile to keep investigating ways to model as precisely as possible lens galaxies for larger samples of systems. Implementing an NFW total density profile could, for example, improve the modelling of galaxy-scale strong lensing systems. On the other hand, stage IV surveys will likely increase the amount of available strong lensing data by several orders of magnitude, possibly mitigating the effect of outliers and thus potentially overcoming issues related to the lens mass model. We finally underline that our model is good at constraining the gravitational slip with fixed screening scales but yields poor constraints on cosmological parameters such as the curvature, dark densities or matter density. This can be attributed to the poor sensitivity of the angular distances ratio to cosmological densities. Time-delay cosmography measurements could however be of interest to constrain the Hubble constant $H_0$.\\

\section{Conclusion}\label{seq: conclusion}

In this work, we used galaxy-scale strong gravitational lensing to constrain deviations from general relativity at the kpc-Mpc scale. A zoo of modified gravity theories have been developed in the past decades to come up with solutions to one or several drawbacks of the concordance model of cosmology $\Lambda$CDM, e.g., to unveil the nature of dark matter and dark energy. We used a phenomenological description of modified gravity theories in the weak-field limit where the gravitational slip parameter $\gamma_{\rm PN}$ captures the deviation from general relativity.\\

Strong lensing data from early-type galaxies with E/S0 morphologies from SLACS and BELLS samples constrain the gravitational slip by measuring the mass of the lens galaxy with two different messengers. On the one hand, using the deflection angle of massless photons in the lens potential and on the other hand by measuring the velocity dispersion of stars and gas in the galactic potential. To do so, a power-law index $\gamma$ models the total density in the lens galaxy. The luminosity density of stars is modelled with a deprojected De Vaucouleurs profile to be compared with the commonly used power-law luminosity density. A degeneracy exists between the gravitational slip $\gamma_{\rm PN}$ and the velocity anisotropy $\beta$; the greater $\beta$, the greater $\gamma_{\rm PN}$. The power-law luminosity density model is sensitive to $\beta$'s prior and can lead to biased estimates of the $\gamma_{\rm PN}$ whereas the De Vaucouleurs profile leads to results quite independent of the prior. A logistic prior on $\beta$ correctly fits recent ETGs data from MaNGA DynPop dynamical modelling. For a constant slip, $\gamma_{\rm PN}=1.14^{+0.22}_{-0.18}$ at the 68\% confidence level for a power-law luminosity density and $\gamma_{\rm PN}=0.90^{+0.18}_{-0.14}$ for a deprojected De Vaucouleurs profile, consistent with GR.\\

Screening effects are ubiquitous in modified gravity theories and appear in high-density regions where general relativity is tested with great precision e.g in the solar system. Inspired by bimetric massive gravity, we parameterise a scale dependent slip by introducing the Vainshtein radius $r_V$ and the Compton wavelength $\lambda_g$ of the theory which represent characteristic scales for screening at small and large scales respectively. We fit the gravitational slip and the power-law index of the total density for various values of the Compton wavelength $\lambda_g$ from the pc to Gpc scales, making the Vainshtein radii of the lens galaxies depend on their mass and $\lambda_g$. We find no statistically significant deviation from GR. Using a De Vaucouleurs deprojected luminosity density, the best-fit is obtained for $\lambda_g \sim 0.2 \text{ Mpc}$ with $\gamma_{\rm PN} = 0.77^{0.25}_{-0.14}$ at the 68\% confidence level. We also find a local minimum for $\lambda_g \sim 100 \text{ Mpc}$ with $\gamma_{\rm PN}=0.56^{0.45}_{-0.35}$. We shed light on the fact that the best-fit obtained for the gravitational slip is correlated with the lens mass model. Having realistic constraints on the lens mass model is a key feature to find good and reliable constraints on the gravitational slip $\gamma_{\rm PN}$ and, a fortiori, any other cosmological parameter of interest. Further investigations on the influence of the lens mass model on cosmological parameters would be worthwhile. Restraining the dataset to fewer samples with excellent knowledge of the lens mass model should reduce the effects associated to outliers and provide more reliable measurements of the gravitational slip. \\

Constraining the deviation from GR is of rising interest with the cosmological surveys to come, e.g Euclid and LSST. Euclid, for example, is expected to provide millions of photometric and spectroscopic galactic observations leading to a sample of strong lenses several order of magnitudes larger than the one employed in this study. It will thus prove of interest, in the years to come, to apply our model to larger samples to see if such an amount of data is able to smooth out effects attributed to outliers. Moreover, strong lensing is not the only way to probe gravity. Fast radio bursts \cite{adiProbingGravitationalSlip2021} or time-delay cosmography \cite{jyotiCosmicTimeSlip2019} are but examples of useful probe to detect gravitational discrepancies from the current concordance model. Time-delay measurement would be of interest since they allow to study the existence of degeneracies between the Hubble constant and the gravitational slip. Let alone our use of strong lensing data, our work have investigated ways to constrain the lens mass model on one hand and to include screening mechanisms on the other hand.

\begin{acknowledgements}
We thank Robert R. Caldwell and Kai Shu for helpful discusssions as well as Yun Chen for sharing his strong lensing data sample. EM acknowledges support from the Swedish Research Council under Dnr VR 2020-03384.
\end{acknowledgements}

\appendix

\section{Empirically identified outliers in our fitted data}

In figure \ref{fig: compare_fit}, we identified persistent outliers between the analysis using a common power-law index $\gamma$ and individual $\gamma$'s for each lens system. Most of those systems were observed in the BELLS survey and studied in Ref. \cite{liStronglensingMeasurementTotalmassdensity2018}. The density slope $\gamma$ fitted is either outside the range $\gamma \in [1.9, 2.1]$ usually obtained, or has unusually large error bars $\Delta \gamma \sim 0.5$.

\begin{table}[h]
    \centering
    \caption{Outliers system identified in Figure \ref{fig: compare_fit}.}
    \begin{tabular}{c c c c c c c}
    \hline 
       Lens name  &  $z_l$ & $z_s$ & $\theta_{E, obs}$ & $\sigma$ & $\Delta \sigma$ & $\gamma$ in \cite{liStronglensingMeasurementTotalmassdensity2018}\\
       \hline
        SDSSJ0237-0641 & 0.4859 & 2.2491 & 0.65 & 290 & 89 & 2.32 $\pm$ 0.27  \\
        SDSSJ0856+2010 & 0.5074 & 2.2335 & 0.98 & 334 & 54 & 2.55 $\pm$ 0.23 \\
        SDSSJ0801+4727 & 0.483 & 1.518 & 0.49 & 98 & 24 & 1.54 $\pm$ 0.27\\
        SDSSJ1234-0241 & 0.49 & 1.016 & 0.53 & 122 & 31 & 1.90 $\pm$ 0.45\\
        SDSSJ0935-0003 & 0.347 & 0.467 & 0.87  & 396 & 35 & -\\
        \hline
    \end{tabular}
    \label{tab: outlier galaxies}
\end{table}

\section{Analytical expression of the velocity dispersion for a power-law luminosity density}

The velocity dispersion in the case of a power-law luminosity density is obtained in Ref. \cite{chenAssessingEffectLens2019} using Jeans equation \eqref{eq: Jeans} to obtain the radial velocity dispersion \eqref{eq: sigma2_r}. The luminosity-weighted average along the line of sight and over the effectives spectroscopic aperture $R_A$ is obtained with equation \eqref{eq: observed VD} and yields with a luminosity density $\nu_{\text{pl}} = \nu_0(r/r_0)^{-\delta}$:

{\tiny
\begin{align}\label{eq: los vd pl}
\begin{split}
    \sigma^2_{\parallel, \rm pl}(\leq R_A) &= \frac{c^2}{2 \sqrt{\pi}}\frac{D_s}{D_{ls}} \theta_{E, \text{GR}} \frac{3-\delta}{(\xi-2\beta)(3-\xi)} \\
    &\times \left[ \frac{\Gamma(\frac{\xi-1}{2})}{\Gamma(\xi/2)}- \beta \frac{\Gamma(\frac{\xi+1}{2})}{\Gamma(\frac{\gamma+2}{2})}\right]\frac{\Gamma(\gamma/2)\Gamma(\delta/2)}{\Gamma(\frac{\gamma-1}{2})\Gamma(\frac{\delta-1}{2})}\left( \frac{\theta_A}{\theta_{E, GR}}\right)^{2-\gamma},
\end{split}
\end{align}
}

where $\xi = \gamma + \delta -2$, $\Gamma$ is the Gamma function and $\theta_A$ is the angular spectroscopic aperture.\\


\bibliography{biblio}

\begin{thebibliography}{46}%
\makeatletter
\providecommand \@ifxundefined [1]{%
 \@ifx{#1\undefined}
}%
\providecommand \@ifnum [1]{%
 \ifnum #1\expandafter \@firstoftwo
 \else \expandafter \@secondoftwo
 \fi
}%
\providecommand \@ifx [1]{%
 \ifx #1\expandafter \@firstoftwo
 \else \expandafter \@secondoftwo
 \fi
}%
\providecommand \natexlab [1]{#1}%
\providecommand \enquote  [1]{``#1''}%
\providecommand \bibnamefont  [1]{#1}%
\providecommand \bibfnamefont [1]{#1}%
\providecommand \citenamefont [1]{#1}%
\providecommand \href@noop [0]{\@secondoftwo}%
\providecommand \href [0]{\begingroup \@sanitize@url \@href}%
\providecommand \@href[1]{\@@startlink{#1}\@@href}%
\providecommand \@@href[1]{\endgroup#1\@@endlink}%
\providecommand \@sanitize@url [0]{\catcode `\\12\catcode `\$12\catcode
  `\&12\catcode `\#12\catcode `\^12\catcode `\_12\catcode `\%12\relax}%
\providecommand \@@startlink[1]{}%
\providecommand \@@endlink[0]{}%
\providecommand \url  [0]{\begingroup\@sanitize@url \@url }%
\providecommand \@url [1]{\endgroup\@href {#1}{\urlprefix }}%
\providecommand \urlprefix  [0]{URL }%
\providecommand \Eprint [0]{\href }%
\providecommand \doibase [0]{https://doi.org/}%
\providecommand \selectlanguage [0]{\@gobble}%
\providecommand \bibinfo  [0]{\@secondoftwo}%
\providecommand \bibfield  [0]{\@secondoftwo}%
\providecommand \translation [1]{[#1]}%
\providecommand \BibitemOpen [0]{}%
\providecommand \bibitemStop [0]{}%
\providecommand \bibitemNoStop [0]{.\EOS\space}%
\providecommand \EOS [0]{\spacefactor3000\relax}%
\providecommand \BibitemShut  [1]{\csname bibitem#1\endcsname}%
\let\auto@bib@innerbib\@empty
\bibitem [{\citenamefont {Blanchard}\ \emph {et~al.}(2020)\citenamefont
  {Blanchard}, \citenamefont {Camera}, \citenamefont {Carbone}, \citenamefont
  {Cardone}, \citenamefont {Casas}, \citenamefont {Clesse}, \citenamefont
  {Ili{\'c}}, \citenamefont {Kilbinger}, \citenamefont {Kitching},
  \citenamefont {Kunz} \emph {et~al.}}]{blanchardEuclidPreparationVII2020}%
  \BibitemOpen
  \bibfield  {author} {\bibinfo {author} {\bibfnamefont {A.}~\bibnamefont
  {Blanchard}}, \bibinfo {author} {\bibfnamefont {S.}~\bibnamefont {Camera}},
  \bibinfo {author} {\bibfnamefont {C.}~\bibnamefont {Carbone}}, \bibinfo
  {author} {\bibfnamefont {V.~F.}\ \bibnamefont {Cardone}}, \bibinfo {author}
  {\bibfnamefont {S.}~\bibnamefont {Casas}}, \bibinfo {author} {\bibfnamefont
  {S.}~\bibnamefont {Clesse}}, \bibinfo {author} {\bibfnamefont
  {S.}~\bibnamefont {Ili{\'c}}}, \bibinfo {author} {\bibfnamefont
  {M.}~\bibnamefont {Kilbinger}}, \bibinfo {author} {\bibfnamefont
  {T.}~\bibnamefont {Kitching}}, \bibinfo {author} {\bibfnamefont
  {M.}~\bibnamefont {Kunz}}, \emph {et~al.},\ }\bibfield  {title} {\bibinfo
  {title} {Euclid preparation - {{VII}}. {{Forecast}} validation for {{Euclid}}
  cosmological probes},\ }\href {https://doi.org/10.1051/0004-6361/202038071}
  {\bibfield  {journal} {\bibinfo  {journal} {Astronomy \& Astrophysics}\
  }\textbf {\bibinfo {volume} {642}},\ \bibinfo {pages} {A191} (\bibinfo {year}
  {2020})}\BibitemShut {NoStop}%
\bibitem [{\citenamefont
  {Weinberg}(1989)}]{weinbergCosmologicalConstantProblem1989}%
  \BibitemOpen
  \bibfield  {author} {\bibinfo {author} {\bibfnamefont {S.}~\bibnamefont
  {Weinberg}},\ }\bibfield  {title} {\bibinfo {title} {The cosmological
  constant problem},\ }\href {https://doi.org/10.1103/RevModPhys.61.1}
  {\bibfield  {journal} {\bibinfo  {journal} {Reviews of Modern Physics}\
  }\textbf {\bibinfo {volume} {61}},\ \bibinfo {pages} {1} (\bibinfo {year}
  {1989})}\BibitemShut {NoStop}%
\bibitem [{\citenamefont {Joyce}\ \emph {et~al.}(2015)\citenamefont {Joyce},
  \citenamefont {Jain}, \citenamefont {Khoury},\ and\ \citenamefont
  {Trodden}}]{joyceCosmologicalStandardModel2015}%
  \BibitemOpen
  \bibfield  {author} {\bibinfo {author} {\bibfnamefont {A.}~\bibnamefont
  {Joyce}}, \bibinfo {author} {\bibfnamefont {B.}~\bibnamefont {Jain}},
  \bibinfo {author} {\bibfnamefont {J.}~\bibnamefont {Khoury}},\ and\ \bibinfo
  {author} {\bibfnamefont {M.}~\bibnamefont {Trodden}},\ }\bibfield  {title}
  {\bibinfo {title} {Beyond the cosmological standard model},\ }\href
  {https://doi.org/10.1016/j.physrep.2014.12.002} {\bibfield  {journal}
  {\bibinfo  {journal} {Physics Reports}\ }\bibinfo {series} {Beyond the
  Cosmological Standard Model},\ \textbf {\bibinfo {volume} {568}},\ \bibinfo
  {pages} {1} (\bibinfo {year} {2015})}\BibitemShut {NoStop}%
\bibitem [{\citenamefont {Shankaranarayanan}\ and\ \citenamefont
  {Johnson}(2022)}]{shankaranarayananModifiedTheoriesGravity2022}%
  \BibitemOpen
  \bibfield  {author} {\bibinfo {author} {\bibfnamefont {S.}~\bibnamefont
  {Shankaranarayanan}}\ and\ \bibinfo {author} {\bibfnamefont {J.~P.}\
  \bibnamefont {Johnson}},\ }\bibfield  {title} {\bibinfo {title} {Modified
  theories of {{Gravity}}: {{Why}}, {{How}} and {{What}}?},\ }\href
  {https://doi.org/10.1007/s10714-022-02927-2} {\bibfield  {journal} {\bibinfo
  {journal} {General Relativity and Gravitation}\ }\textbf {\bibinfo {volume}
  {54}},\ \bibinfo {pages} {44} (\bibinfo {year} {2022})},\ \Eprint
  {https://arxiv.org/abs/2204.06533} {arxiv:2204.06533 [astro-ph,
  physics:gr-qc, physics:hep-th]} \BibitemShut {NoStop}%
\bibitem [{\citenamefont {Schlamminger}\ \emph {et~al.}(2008)\citenamefont
  {Schlamminger}, \citenamefont {Choi}, \citenamefont {Wagner}, \citenamefont
  {Gundlach},\ and\ \citenamefont
  {Adelberger}}]{schlammingerTestEquivalencePrinciple2008}%
  \BibitemOpen
  \bibfield  {author} {\bibinfo {author} {\bibfnamefont {S.}~\bibnamefont
  {Schlamminger}}, \bibinfo {author} {\bibfnamefont {K.-Y.}\ \bibnamefont
  {Choi}}, \bibinfo {author} {\bibfnamefont {T.~A.}\ \bibnamefont {Wagner}},
  \bibinfo {author} {\bibfnamefont {J.~H.}\ \bibnamefont {Gundlach}},\ and\
  \bibinfo {author} {\bibfnamefont {E.~G.}\ \bibnamefont {Adelberger}},\
  }\bibfield  {title} {\bibinfo {title} {Test of the {{Equivalence Principle
  Using}} a {{Rotating Torsion Balance}}},\ }\href
  {https://doi.org/10.1103/PhysRevLett.100.041101} {\bibfield  {journal}
  {\bibinfo  {journal} {Physical Review Letters}\ }\textbf {\bibinfo {volume}
  {100}},\ \bibinfo {pages} {041101} (\bibinfo {year} {2008})},\ \Eprint
  {https://arxiv.org/abs/0712.0607} {arxiv:0712.0607 [gr-qc]} \BibitemShut
  {NoStop}%
\bibitem [{\citenamefont {Shapiro}(1964)}]{shapiroFourthTestGeneral1964}%
  \BibitemOpen
  \bibfield  {author} {\bibinfo {author} {\bibfnamefont {I.~I.}\ \bibnamefont
  {Shapiro}},\ }\bibfield  {title} {\bibinfo {title} {Fourth {{Test}} of
  {{General Relativity}}},\ }\href {https://doi.org/10.1103/PhysRevLett.13.789}
  {\bibfield  {journal} {\bibinfo  {journal} {Physical Review Letters}\
  }\textbf {\bibinfo {volume} {13}},\ \bibinfo {pages} {789} (\bibinfo {year}
  {1964})}\BibitemShut {NoStop}%
\bibitem [{\citenamefont {Pound}\ and\ \citenamefont
  {Rebka}(1960)}]{poundApparentWeightPhotons1960}%
  \BibitemOpen
  \bibfield  {author} {\bibinfo {author} {\bibfnamefont {R.~V.}\ \bibnamefont
  {Pound}}\ and\ \bibinfo {author} {\bibfnamefont {G.~A.}\ \bibnamefont
  {Rebka}},\ }\bibfield  {title} {\bibinfo {title} {Apparent {{Weight}} of
  {{Photons}}},\ }\href {https://doi.org/10.1103/PhysRevLett.4.337} {\bibfield
  {journal} {\bibinfo  {journal} {Physical Review Letters}\ }\textbf {\bibinfo
  {volume} {4}},\ \bibinfo {pages} {337} (\bibinfo {year} {1960})}\BibitemShut
  {NoStop}%
\bibitem [{\citenamefont {Koyama}(2016)}]{koyamaCosmologicalTestsModified2016}%
  \BibitemOpen
  \bibfield  {author} {\bibinfo {author} {\bibfnamefont {K.}~\bibnamefont
  {Koyama}},\ }\bibfield  {title} {\bibinfo {title} {Cosmological {{Tests}} of
  {{Modified Gravity}}},\ }\href
  {https://doi.org/10.1088/0034-4885/79/4/046902} {\bibfield  {journal}
  {\bibinfo  {journal} {Reports on Progress in Physics}\ }\textbf {\bibinfo
  {volume} {79}},\ \bibinfo {pages} {046902} (\bibinfo {year} {2016})},\
  \Eprint {https://arxiv.org/abs/1504.04623} {arxiv:1504.04623 [astro-ph,
  physics:gr-qc, physics:hep-ph, physics:hep-th]} \BibitemShut {NoStop}%
\bibitem [{\citenamefont {Thorne}\ and\ \citenamefont
  {Will}(1971)}]{thorneTheoreticalFrameworksTesting1971}%
  \BibitemOpen
  \bibfield  {author} {\bibinfo {author} {\bibfnamefont {K.~S.}\ \bibnamefont
  {Thorne}}\ and\ \bibinfo {author} {\bibfnamefont {C.~M.}\ \bibnamefont
  {Will}},\ }\bibfield  {title} {\bibinfo {title} {Theoretical {{Frameworks}}
  for {{Testing Relativistic Gravity}}. {{I}}. {{Foundations}}},\ }\href
  {https://doi.org/10.1086/150803} {\bibfield  {journal} {\bibinfo  {journal}
  {The Astrophysical Journal}\ }\textbf {\bibinfo {volume} {163}},\ \bibinfo
  {pages} {595} (\bibinfo {year} {1971})}\BibitemShut {NoStop}%
\bibitem [{\citenamefont {Cao}\ \emph {et~al.}(2015)\citenamefont {Cao},
  \citenamefont {Biesiada}, \citenamefont {Gavazzi}, \citenamefont
  {Pi{\'o}rkowska},\ and\ \citenamefont
  {Zhu}}]{caoCOSMOLOGYSTRONGLENSINGSYSTEMS2015}%
  \BibitemOpen
  \bibfield  {author} {\bibinfo {author} {\bibfnamefont {S.}~\bibnamefont
  {Cao}}, \bibinfo {author} {\bibfnamefont {M.}~\bibnamefont {Biesiada}},
  \bibinfo {author} {\bibfnamefont {R.}~\bibnamefont {Gavazzi}}, \bibinfo
  {author} {\bibfnamefont {A.}~\bibnamefont {Pi{\'o}rkowska}},\ and\ \bibinfo
  {author} {\bibfnamefont {Z.-H.}\ \bibnamefont {Zhu}},\ }\bibfield  {title}
  {\bibinfo {title} {{{Cosmology with strong-lensing systems}}},\ }\href
  {https://doi.org/10.1088/0004-637X/806/2/185} {\bibfield  {journal} {\bibinfo
   {journal} {The Astrophysical Journal}\ }\textbf {\bibinfo {volume} {806}},\
  \bibinfo {pages} {185} (\bibinfo {year} {2015})}\BibitemShut {NoStop}%
\bibitem [{\citenamefont {Amante}\ \emph {et~al.}(2020)\citenamefont {Amante},
  \citenamefont {Maga{\~n}a}, \citenamefont {Motta}, \citenamefont
  {{Garc{\'i}a-Aspeitia}},\ and\ \citenamefont
  {Verdugo}}]{amanteTestingDarkEnergy2020}%
  \BibitemOpen
  \bibfield  {author} {\bibinfo {author} {\bibfnamefont {M.~H.}\ \bibnamefont
  {Amante}}, \bibinfo {author} {\bibfnamefont {J.}~\bibnamefont {Maga{\~n}a}},
  \bibinfo {author} {\bibfnamefont {V.}~\bibnamefont {Motta}}, \bibinfo
  {author} {\bibfnamefont {M.~A.}\ \bibnamefont {{Garc{\'i}a-Aspeitia}}},\ and\
  \bibinfo {author} {\bibfnamefont {T.}~\bibnamefont {Verdugo}},\ }\bibfield
  {title} {\bibinfo {title} {Testing dark energy models with a new sample of
  strong-lensing systems},\ }\href {https://doi.org/10.1093/mnras/staa2760}
  {\bibfield  {journal} {\bibinfo  {journal} {Monthly Notices of the Royal
  Astronomical Society}\ }\textbf {\bibinfo {volume} {498}},\ \bibinfo {pages}
  {6013} (\bibinfo {year} {2020})},\ \Eprint {https://arxiv.org/abs/1906.04107}
  {arxiv:1906.04107 [astro-ph, physics:gr-qc]} \BibitemShut {NoStop}%
\bibitem [{\citenamefont {Birrer}\ \emph {et~al.}(2019)\citenamefont {Birrer},
  \citenamefont {Treu}, \citenamefont {Rusu}, \citenamefont {Bonvin},
  \citenamefont {Fassnacht}, \citenamefont {Chan}, \citenamefont {Agnello},
  \citenamefont {Shajib}, \citenamefont {Chen}, \citenamefont {Auger} \emph
  {et~al.}}]{birrerH0LiCOWIXCosmographic2019}%
  \BibitemOpen
  \bibfield  {author} {\bibinfo {author} {\bibfnamefont {S.}~\bibnamefont
  {Birrer}}, \bibinfo {author} {\bibfnamefont {T.}~\bibnamefont {Treu}},
  \bibinfo {author} {\bibfnamefont {C.~E.}\ \bibnamefont {Rusu}}, \bibinfo
  {author} {\bibfnamefont {V.}~\bibnamefont {Bonvin}}, \bibinfo {author}
  {\bibfnamefont {C.~D.}\ \bibnamefont {Fassnacht}}, \bibinfo {author}
  {\bibfnamefont {J.~H.~H.}\ \bibnamefont {Chan}}, \bibinfo {author}
  {\bibfnamefont {A.}~\bibnamefont {Agnello}}, \bibinfo {author} {\bibfnamefont
  {A.~J.}\ \bibnamefont {Shajib}}, \bibinfo {author} {\bibfnamefont {G.~C.~F.}\
  \bibnamefont {Chen}}, \bibinfo {author} {\bibfnamefont {M.}~\bibnamefont
  {Auger}}, \emph {et~al.},\ }\bibfield  {title} {\bibinfo {title} {{{H0LiCOW}}
  - {{IX}}. {{Cosmographic}} analysis of the doubly imaged quasar {{SDSS}}
  1206+4332 and a new measurement of the {{Hubble}} constant},\ }\href
  {https://doi.org/10.1093/mnras/stz200} {\bibfield  {journal} {\bibinfo
  {journal} {Monthly Notices of the Royal Astronomical Society}\ }\textbf
  {\bibinfo {volume} {484}},\ \bibinfo {pages} {4726} (\bibinfo {year}
  {2019})}\BibitemShut {NoStop}%
\bibitem [{\citenamefont {Wong}\ \emph {et~al.}(2020)\citenamefont {Wong},
  \citenamefont {Suyu}, \citenamefont {Chen}, \citenamefont {Rusu},
  \citenamefont {Millon}, \citenamefont {Sluse}, \citenamefont {Bonvin},
  \citenamefont {Fassnacht}, \citenamefont {Taubenberger}, \citenamefont
  {Auger} \emph {et~al.}}]{wongH0LiCOWXIIICent2020}%
  \BibitemOpen
  \bibfield  {author} {\bibinfo {author} {\bibfnamefont {K.~C.}\ \bibnamefont
  {Wong}}, \bibinfo {author} {\bibfnamefont {S.~H.}\ \bibnamefont {Suyu}},
  \bibinfo {author} {\bibfnamefont {G.~C.~F.}\ \bibnamefont {Chen}}, \bibinfo
  {author} {\bibfnamefont {C.~E.}\ \bibnamefont {Rusu}}, \bibinfo {author}
  {\bibfnamefont {M.}~\bibnamefont {Millon}}, \bibinfo {author} {\bibfnamefont
  {D.}~\bibnamefont {Sluse}}, \bibinfo {author} {\bibfnamefont
  {V.}~\bibnamefont {Bonvin}}, \bibinfo {author} {\bibfnamefont {C.~D.}\
  \bibnamefont {Fassnacht}}, \bibinfo {author} {\bibfnamefont {S.}~\bibnamefont
  {Taubenberger}}, \bibinfo {author} {\bibfnamefont {M.~W.}\ \bibnamefont
  {Auger}}, \emph {et~al.},\ }\bibfield  {title} {\bibinfo {title} {{{H0LiCOW}}
  - {{XIII}}. {{A}} 2.4 per cent measurement of {{H0}} from lensed quasars:
  5.3{$\sigma$} tension between early- and late-{{Universe}} probes},\ }\href
  {https://doi.org/10.1093/mnras/stz3094} {\bibfield  {journal} {\bibinfo
  {journal} {Monthly Notices of the Royal Astronomical Society}\ }\textbf
  {\bibinfo {volume} {498}},\ \bibinfo {pages} {1420} (\bibinfo {year}
  {2020})}\BibitemShut {NoStop}%
\bibitem [{\citenamefont {Liu}\ \emph {et~al.}(2020)\citenamefont {Liu},
  \citenamefont {Cao}, \citenamefont {Zhang}, \citenamefont {Biesiada},
  \citenamefont {Liu},\ and\ \citenamefont
  {Lian}}]{liuTestingCosmicCurvature2020}%
  \BibitemOpen
  \bibfield  {author} {\bibinfo {author} {\bibfnamefont {T.}~\bibnamefont
  {Liu}}, \bibinfo {author} {\bibfnamefont {S.}~\bibnamefont {Cao}}, \bibinfo
  {author} {\bibfnamefont {J.}~\bibnamefont {Zhang}}, \bibinfo {author}
  {\bibfnamefont {M.}~\bibnamefont {Biesiada}}, \bibinfo {author}
  {\bibfnamefont {Y.}~\bibnamefont {Liu}},\ and\ \bibinfo {author}
  {\bibfnamefont {Y.}~\bibnamefont {Lian}},\ }\bibfield  {title} {\bibinfo
  {title} {Testing the cosmic curvature at high redshifts: The combination of
  {{LSST}} strong lensing systems and quasars as new standard candles},\ }\href
  {https://doi.org/10.1093/mnras/staa1539} {\bibfield  {journal} {\bibinfo
  {journal} {Monthly Notices of the Royal Astronomical Society}\ }\textbf
  {\bibinfo {volume} {496}},\ \bibinfo {pages} {708} (\bibinfo {year}
  {2020})}\BibitemShut {NoStop}%
\bibitem [{\citenamefont {Cao}\ \emph {et~al.}(2016)\citenamefont {Cao},
  \citenamefont {Biesiada}, \citenamefont {Yao},\ and\ \citenamefont
  {Zhu}}]{caoLimitsPowerlawMass2016}%
  \BibitemOpen
  \bibfield  {author} {\bibinfo {author} {\bibfnamefont {S.}~\bibnamefont
  {Cao}}, \bibinfo {author} {\bibfnamefont {M.}~\bibnamefont {Biesiada}},
  \bibinfo {author} {\bibfnamefont {M.}~\bibnamefont {Yao}},\ and\ \bibinfo
  {author} {\bibfnamefont {Z.-H.}\ \bibnamefont {Zhu}},\ }\bibfield  {title}
  {\bibinfo {title} {Limits on the power-law mass and luminosity density
  profiles of elliptical galaxies from gravitational lensing systems},\ }\href
  {https://doi.org/10.1093/mnras/stw932} {\bibfield  {journal} {\bibinfo
  {journal} {Monthly Notices of the Royal Astronomical Society}\ }\textbf
  {\bibinfo {volume} {461}},\ \bibinfo {pages} {2192} (\bibinfo {year}
  {2016})}\BibitemShut {NoStop}%
\bibitem [{\citenamefont {Chen}\ \emph {et~al.}(2019)\citenamefont {Chen},
  \citenamefont {Li}, \citenamefont {Shu},\ and\ \citenamefont
  {Cao}}]{chenAssessingEffectLens2019}%
  \BibitemOpen
  \bibfield  {author} {\bibinfo {author} {\bibfnamefont {Y.}~\bibnamefont
  {Chen}}, \bibinfo {author} {\bibfnamefont {R.}~\bibnamefont {Li}}, \bibinfo
  {author} {\bibfnamefont {Y.}~\bibnamefont {Shu}},\ and\ \bibinfo {author}
  {\bibfnamefont {X.}~\bibnamefont {Cao}},\ }\bibfield  {title} {\bibinfo
  {title} {Assessing the effect of lens mass model in cosmological application
  with updated galaxy-scale strong gravitational lensing sample},\ }\href
  {https://doi.org/10.1093/mnras/stz1902} {\bibfield  {journal} {\bibinfo
  {journal} {Monthly Notices of the Royal Astronomical Society}\ }\textbf
  {\bibinfo {volume} {488}},\ \bibinfo {pages} {3745} (\bibinfo {year}
  {2019})},\ \Eprint {https://arxiv.org/abs/1809.09845} {arxiv:1809.09845
  [astro-ph]} \BibitemShut {NoStop}%
\bibitem [{\citenamefont {Cao}\ \emph {et~al.}(2017)\citenamefont {Cao},
  \citenamefont {Li}, \citenamefont {Biesiada}, \citenamefont {Xu},
  \citenamefont {Cai},\ and\ \citenamefont
  {Zhu}}]{caoTESTPARAMETRIZEDPOSTNEWTONIAN2017}%
  \BibitemOpen
  \bibfield  {author} {\bibinfo {author} {\bibfnamefont {S.}~\bibnamefont
  {Cao}}, \bibinfo {author} {\bibfnamefont {X.}~\bibnamefont {Li}}, \bibinfo
  {author} {\bibfnamefont {M.}~\bibnamefont {Biesiada}}, \bibinfo {author}
  {\bibfnamefont {T.}~\bibnamefont {Xu}}, \bibinfo {author} {\bibfnamefont
  {Y.}~\bibnamefont {Cai}},\ and\ \bibinfo {author} {\bibfnamefont {Z.-H.}\
  \bibnamefont {Zhu}},\ }\bibfield  {title} {\bibinfo {title} {{{Test of
  parametrized Post-Newtonian gravity with galaxy-scale Strong Lensing
  systems}}},\ }\href {https://doi.org/10.3847/1538-4357/835/1/92} {\bibfield
  {journal} {\bibinfo  {journal} {The Astrophysical Journal}\ }\textbf
  {\bibinfo {volume} {835}},\ \bibinfo {pages} {92} (\bibinfo {year}
  {2017})}\BibitemShut {NoStop}%
\bibitem [{\citenamefont {Wei}\ \emph {et~al.}(2022)\citenamefont {Wei},
  \citenamefont {Chen}, \citenamefont {Cao},\ and\ \citenamefont
  {Wu}}]{weiDirectEstimatePostNewtonian2022}%
  \BibitemOpen
  \bibfield  {author} {\bibinfo {author} {\bibfnamefont {J.-J.}\ \bibnamefont
  {Wei}}, \bibinfo {author} {\bibfnamefont {Y.}~\bibnamefont {Chen}}, \bibinfo
  {author} {\bibfnamefont {S.}~\bibnamefont {Cao}},\ and\ \bibinfo {author}
  {\bibfnamefont {X.-F.}\ \bibnamefont {Wu}},\ }\bibfield  {title} {\bibinfo
  {title} {Direct {{Estimate}} of the {{Post-Newtonian Parameter}} and {{Cosmic
  Curvature}} from {{Galaxy-scale Strong Gravitational Lensing}}},\ }\href
  {https://doi.org/10.3847/2041-8213/ac551e} {\bibfield  {journal} {\bibinfo
  {journal} {The Astrophysical Journal Letters}\ }\textbf {\bibinfo {volume}
  {927}},\ \bibinfo {pages} {L1} (\bibinfo {year} {2022})},\ \Eprint
  {https://arxiv.org/abs/2202.07860} {arxiv:2202.07860 [astro-ph,
  physics:gr-qc, physics:hep-ph]} \BibitemShut {NoStop}%
\bibitem [{\citenamefont {Liu}\ \emph {et~al.}(2022)\citenamefont {Liu},
  \citenamefont {Li}, \citenamefont {Qi},\ and\ \citenamefont
  {Zhang}}]{liuGalaxyscaleTestGeneral2022}%
  \BibitemOpen
  \bibfield  {author} {\bibinfo {author} {\bibfnamefont {X.-H.}\ \bibnamefont
  {Liu}}, \bibinfo {author} {\bibfnamefont {Z.-H.}\ \bibnamefont {Li}},
  \bibinfo {author} {\bibfnamefont {J.-Z.}\ \bibnamefont {Qi}},\ and\ \bibinfo
  {author} {\bibfnamefont {X.}~\bibnamefont {Zhang}},\ }\bibfield  {title}
  {\bibinfo {title} {Galaxy-scale {{Test}} of {{General Relativity}} with
  {{Strong Gravitational Lensing}}},\ }\href
  {https://doi.org/10.3847/1538-4357/ac4c3b} {\bibfield  {journal} {\bibinfo
  {journal} {The Astrophysical Journal}\ }\textbf {\bibinfo {volume} {927}},\
  \bibinfo {pages} {28} (\bibinfo {year} {2022})}\BibitemShut {NoStop}%
\bibitem [{\citenamefont {Lian}\ \emph {et~al.}(2022)\citenamefont {Lian},
  \citenamefont {Cao}, \citenamefont {Liu}, \citenamefont {Biesiada},\ and\
  \citenamefont {Zhu}}]{lianDirectTestsGeneral2022}%
  \BibitemOpen
  \bibfield  {author} {\bibinfo {author} {\bibfnamefont {Y.}~\bibnamefont
  {Lian}}, \bibinfo {author} {\bibfnamefont {S.}~\bibnamefont {Cao}}, \bibinfo
  {author} {\bibfnamefont {T.}~\bibnamefont {Liu}}, \bibinfo {author}
  {\bibfnamefont {M.}~\bibnamefont {Biesiada}},\ and\ \bibinfo {author}
  {\bibfnamefont {Z.-H.}\ \bibnamefont {Zhu}},\ }\bibfield  {title} {\bibinfo
  {title} {Direct tests of {{General Relativity}} under screening effect with
  galaxy-scale strong lensing systems},\ }\href
  {https://doi.org/10.3847/1538-4357/ac9d36} {\bibfield  {journal} {\bibinfo
  {journal} {The Astrophysical Journal}\ }\textbf {\bibinfo {volume} {941}},\
  \bibinfo {pages} {16} (\bibinfo {year} {2022})},\ \Eprint
  {https://arxiv.org/abs/2210.16752} {arxiv:2210.16752 [astro-ph,
  physics:gr-qc]} \BibitemShut {NoStop}%
\bibitem [{\citenamefont {Jyoti}\ \emph {et~al.}(2019)\citenamefont {Jyoti},
  \citenamefont {Mu{\~n}oz}, \citenamefont {Caldwell},\ and\ \citenamefont
  {Kamionkowski}}]{jyotiCosmicTimeSlip2019}%
  \BibitemOpen
  \bibfield  {author} {\bibinfo {author} {\bibfnamefont {D.}~\bibnamefont
  {Jyoti}}, \bibinfo {author} {\bibfnamefont {J.~B.}\ \bibnamefont
  {Mu{\~n}oz}}, \bibinfo {author} {\bibfnamefont {R.~R.}\ \bibnamefont
  {Caldwell}},\ and\ \bibinfo {author} {\bibfnamefont {M.}~\bibnamefont
  {Kamionkowski}},\ }\bibfield  {title} {\bibinfo {title} {Cosmic time slip:
  {{Testing}} gravity on supergalactic scales with strong-lensing time
  delays},\ }\href {https://doi.org/10.1103/PhysRevD.100.043031} {\bibfield
  {journal} {\bibinfo  {journal} {Physical Review D}\ }\textbf {\bibinfo
  {volume} {100}},\ \bibinfo {pages} {043031} (\bibinfo {year}
  {2019})}\BibitemShut {NoStop}%
\bibitem [{\citenamefont {Adi}\ and\ \citenamefont
  {Kovetz}(2021)}]{adiProbingGravitationalSlip2021}%
  \BibitemOpen
  \bibfield  {author} {\bibinfo {author} {\bibfnamefont {T.}~\bibnamefont
  {Adi}}\ and\ \bibinfo {author} {\bibfnamefont {E.~D.}\ \bibnamefont
  {Kovetz}},\ }\bibfield  {title} {\bibinfo {title} {Probing gravitational slip
  with strongly lensed fast radio bursts},\ }\href
  {https://doi.org/10.1103/PhysRevD.104.103515} {\bibfield  {journal} {\bibinfo
   {journal} {Physical Review D}\ }\textbf {\bibinfo {volume} {104}},\ \bibinfo
  {pages} {103515} (\bibinfo {year} {2021})}\BibitemShut {NoStop}%
\bibitem [{\citenamefont {Schneider}\ \emph {et~al.}(2006)\citenamefont
  {Schneider}, \citenamefont {Kochanek},\ and\ \citenamefont
  {Wambsganss}}]{schneiderGravitationalLensingStrong2006}%
  \BibitemOpen
  \bibfield  {author} {\bibinfo {author} {\bibfnamefont {P.}~\bibnamefont
  {Schneider}}, \bibinfo {author} {\bibfnamefont {C.~S.}\ \bibnamefont
  {Kochanek}},\ and\ \bibinfo {author} {\bibfnamefont {J.}~\bibnamefont
  {Wambsganss}},\ }\href {https://doi.org/10.1007/978-3-540-30310-7} {\emph
  {\bibinfo {title} {Gravitational {{Lensing}}: {{Strong}}, {{Weak}} and
  {{Micro}}}}},\ edited by\ \bibinfo {editor} {\bibfnamefont {G.}~\bibnamefont
  {Meylan}}\ and\ \bibinfo {editor} {\bibfnamefont {P.}~\bibnamefont
  {Jetzer}},\ \bibinfo {series} {Saas-{{Fee Advanced Courses}}}, Vol.~\bibinfo
  {volume} {33}\ (\bibinfo  {publisher} {{Springer}},\ \bibinfo {address}
  {{Berlin, Heidelberg}},\ \bibinfo {year} {2006})\BibitemShut {NoStop}%
\bibitem [{\citenamefont {Binney}\ and\ \citenamefont
  {Tremaine}(2008)}]{binneyGalacticDynamicsSecond2008}%
  \BibitemOpen
  \bibfield  {author} {\bibinfo {author} {\bibfnamefont {J.}~\bibnamefont
  {Binney}}\ and\ \bibinfo {author} {\bibfnamefont {S.}~\bibnamefont
  {Tremaine}},\ }\href {https://doi.org/10.2307/j.ctvc778ff} {\emph {\bibinfo
  {title} {Galactic {{Dynamics}}: {{Second Edition}}}}},\ \bibinfo {edition}
  {rev - revised, 2}\ ed.\ (\bibinfo  {publisher} {{Princeton University
  Press}},\ \bibinfo {year} {2008})\ \Eprint
  {https://arxiv.org/abs/j.ctvc778ff} {j.ctvc778ff} \BibitemShut {NoStop}%
\bibitem [{\citenamefont {Mellier}\ and\ \citenamefont
  {Mathez}(1987)}]{mellierDeprojectionVaucouleursExp1987}%
  \BibitemOpen
  \bibfield  {author} {\bibinfo {author} {\bibfnamefont {Y.}~\bibnamefont
  {Mellier}}\ and\ \bibinfo {author} {\bibfnamefont {G.}~\bibnamefont
  {Mathez}},\ }\bibfield  {title} {\bibinfo {title} {Deprojection of the de
  {{Vaucouleurs R}} exp 1/4 brightness profile},\ }\href@noop {} {\bibfield
  {journal} {\bibinfo  {journal} {Astronomy and Astrophysics}\ }\textbf
  {\bibinfo {volume} {175}},\ \bibinfo {pages} {1} (\bibinfo {year}
  {1987})}\BibitemShut {NoStop}%
\bibitem [{\citenamefont {Ma}\ and\ \citenamefont
  {Bertschinger}(1995)}]{maCosmologicalPerturbationTheory1995}%
  \BibitemOpen
  \bibfield  {author} {\bibinfo {author} {\bibfnamefont {C.-P.}\ \bibnamefont
  {Ma}}\ and\ \bibinfo {author} {\bibfnamefont {E.}~\bibnamefont
  {Bertschinger}},\ }\bibfield  {title} {\bibinfo {title} {Cosmological
  {{Perturbation Theory}} in the {{Synchronous}} and {{Conformal Newtonian
  Gauges}}},\ }\href {https://doi.org/10.1086/176550} {\bibfield  {journal}
  {\bibinfo  {journal} {The Astrophysical Journal}\ }\textbf {\bibinfo {volume}
  {455}},\ \bibinfo {pages} {7} (\bibinfo {year} {1995})}\BibitemShut {NoStop}%
\bibitem [{\citenamefont {Sotiriou}\ and\ \citenamefont
  {Faraoni}(2010)}]{sotiriouTheoriesGravity2010}%
  \BibitemOpen
  \bibfield  {author} {\bibinfo {author} {\bibfnamefont {T.~P.}\ \bibnamefont
  {Sotiriou}}\ and\ \bibinfo {author} {\bibfnamefont {V.}~\bibnamefont
  {Faraoni}},\ }\bibfield  {title} {\bibinfo {title} {F({{R}}) {{Theories Of
  Gravity}}},\ }\href {https://doi.org/10.1103/RevModPhys.82.451} {\bibfield
  {journal} {\bibinfo  {journal} {Reviews of Modern Physics}\ }\textbf
  {\bibinfo {volume} {82}},\ \bibinfo {pages} {451} (\bibinfo {year} {2010})},\
  \Eprint {https://arxiv.org/abs/0805.1726} {arxiv:0805.1726 [astro-ph,
  physics:gr-qc, physics:hep-th]} \BibitemShut {NoStop}%
\bibitem [{\citenamefont {{Schmidt-May}}\ and\ \citenamefont {{von
  Strauss}}(2016)}]{schmidt-mayRecentDevelopmentsBimetric2016}%
  \BibitemOpen
  \bibfield  {author} {\bibinfo {author} {\bibfnamefont {A.}~\bibnamefont
  {{Schmidt-May}}}\ and\ \bibinfo {author} {\bibfnamefont {M.}~\bibnamefont
  {{von Strauss}}},\ }\bibfield  {title} {\bibinfo {title} {Recent developments
  in bimetric theory},\ }\href {https://doi.org/10.1088/1751-8113/49/18/183001}
  {\bibfield  {journal} {\bibinfo  {journal} {Journal of Physics A:
  Mathematical and Theoretical}\ }\textbf {\bibinfo {volume} {49}},\ \bibinfo
  {pages} {183001} (\bibinfo {year} {2016})},\ \Eprint
  {https://arxiv.org/abs/1512.00021} {arxiv:1512.00021 [gr-qc, physics:hep-th]}
  \BibitemShut {NoStop}%
\bibitem [{\citenamefont {Enander}\ and\ \citenamefont
  {Mortsell}(2013)}]{enanderStrongLensingConstraints2013}%
  \BibitemOpen
  \bibfield  {author} {\bibinfo {author} {\bibfnamefont {J.}~\bibnamefont
  {Enander}}\ and\ \bibinfo {author} {\bibfnamefont {E.}~\bibnamefont
  {Mortsell}},\ }\bibfield  {title} {\bibinfo {title} {Strong lensing
  constraints on bimetric massive gravity},\ }\href
  {https://doi.org/10.1007/JHEP10(2013)031} {\bibfield  {journal} {\bibinfo
  {journal} {Journal of High Energy Physics}\ }\textbf {\bibinfo {volume}
  {2013}},\ \bibinfo {pages} {31} (\bibinfo {year} {2013})},\ \Eprint
  {https://arxiv.org/abs/1306.1086} {arxiv:1306.1086 [astro-ph, physics:gr-qc,
  physics:hep-th]} \BibitemShut {NoStop}%
\bibitem [{\citenamefont {Schwab}\ \emph {et~al.}(2010)\citenamefont {Schwab},
  \citenamefont {Bolton},\ and\ \citenamefont
  {Rappaport}}]{schwabGalaxyScaleStrongLensingTests2010}%
  \BibitemOpen
  \bibfield  {author} {\bibinfo {author} {\bibfnamefont {J.}~\bibnamefont
  {Schwab}}, \bibinfo {author} {\bibfnamefont {A.~S.}\ \bibnamefont {Bolton}},\
  and\ \bibinfo {author} {\bibfnamefont {S.~A.}\ \bibnamefont {Rappaport}},\
  }\bibfield  {title} {\bibinfo {title} {Galaxy-{{Scale Strong-Lensing Tests}}
  of {{Gravity}} and {{Geometric Cosmology}}: {{Constraints}} and {{Systematic
  Limitations}}},\ }\href {https://doi.org/10.1088/0004-637X/708/1/750}
  {\bibfield  {journal} {\bibinfo  {journal} {The Astrophysical Journal}\
  }\textbf {\bibinfo {volume} {708}},\ \bibinfo {pages} {750} (\bibinfo {year}
  {2010})}\BibitemShut {NoStop}%
\bibitem [{\citenamefont {Babichev}\ and\ \citenamefont
  {Crisostomi}(2013)}]{babichevRestoringGeneralRelativity2013}%
  \BibitemOpen
  \bibfield  {author} {\bibinfo {author} {\bibfnamefont {E.}~\bibnamefont
  {Babichev}}\ and\ \bibinfo {author} {\bibfnamefont {M.}~\bibnamefont
  {Crisostomi}},\ }\bibfield  {title} {\bibinfo {title} {Restoring {{General
  Relativity}} in massive bi-gravity theory},\ }\href
  {https://doi.org/10.1103/PhysRevD.88.084002} {\bibfield  {journal} {\bibinfo
  {journal} {Physical Review D}\ }\textbf {\bibinfo {volume} {88}},\ \bibinfo
  {pages} {084002} (\bibinfo {year} {2013})},\ \Eprint
  {https://arxiv.org/abs/1307.3640} {arxiv:1307.3640 [gr-qc, physics:hep-th]}
  \BibitemShut {NoStop}%
\bibitem [{\citenamefont {Bolton}\ \emph {et~al.}(2008)\citenamefont {Bolton},
  \citenamefont {Burles}, \citenamefont {Koopmans}, \citenamefont {Treu},
  \citenamefont {Gavazzi}, \citenamefont {Moustakas}, \citenamefont {Wayth},\
  and\ \citenamefont {Schlegel}}]{boltonSloanLensACS2008}%
  \BibitemOpen
  \bibfield  {author} {\bibinfo {author} {\bibfnamefont {A.~S.}\ \bibnamefont
  {Bolton}}, \bibinfo {author} {\bibfnamefont {S.}~\bibnamefont {Burles}},
  \bibinfo {author} {\bibfnamefont {L.~V.~E.}\ \bibnamefont {Koopmans}},
  \bibinfo {author} {\bibfnamefont {T.}~\bibnamefont {Treu}}, \bibinfo {author}
  {\bibfnamefont {R.}~\bibnamefont {Gavazzi}}, \bibinfo {author} {\bibfnamefont
  {L.~A.}\ \bibnamefont {Moustakas}}, \bibinfo {author} {\bibfnamefont
  {R.}~\bibnamefont {Wayth}},\ and\ \bibinfo {author} {\bibfnamefont {D.~J.}\
  \bibnamefont {Schlegel}},\ }\bibfield  {title} {\bibinfo {title} {The {{Sloan
  Lens ACS Survey}}. {{V}}. {{The Full ACS Strong-Lens Sample}}},\ }\href
  {https://doi.org/10.1086/589327} {\bibfield  {journal} {\bibinfo  {journal}
  {The Astrophysical Journal}\ }\textbf {\bibinfo {volume} {682}},\ \bibinfo
  {pages} {964} (\bibinfo {year} {2008})},\ \Eprint
  {https://arxiv.org/abs/0805.1931} {arxiv:0805.1931 [astro-ph]} \BibitemShut
  {NoStop}%
\bibitem [{\citenamefont {Shu}\ \emph {et~al.}(2017)\citenamefont {Shu},
  \citenamefont {Brownstein}, \citenamefont {Bolton}, \citenamefont {Koopmans},
  \citenamefont {Treu}, \citenamefont {{Montero-Dorta}}, \citenamefont {Auger},
  \citenamefont {Czoske}, \citenamefont {Gavazzi}, \citenamefont {Marshall}
  \emph {et~al.}}]{shuSloanLensACS2017}%
  \BibitemOpen
  \bibfield  {author} {\bibinfo {author} {\bibfnamefont {Y.}~\bibnamefont
  {Shu}}, \bibinfo {author} {\bibfnamefont {J.~R.}\ \bibnamefont {Brownstein}},
  \bibinfo {author} {\bibfnamefont {A.~S.}\ \bibnamefont {Bolton}}, \bibinfo
  {author} {\bibfnamefont {L.~V.~E.}\ \bibnamefont {Koopmans}}, \bibinfo
  {author} {\bibfnamefont {T.}~\bibnamefont {Treu}}, \bibinfo {author}
  {\bibfnamefont {A.~D.}\ \bibnamefont {{Montero-Dorta}}}, \bibinfo {author}
  {\bibfnamefont {M.~W.}\ \bibnamefont {Auger}}, \bibinfo {author}
  {\bibfnamefont {O.}~\bibnamefont {Czoske}}, \bibinfo {author} {\bibfnamefont
  {R.}~\bibnamefont {Gavazzi}}, \bibinfo {author} {\bibfnamefont {P.~J.}\
  \bibnamefont {Marshall}}, \emph {et~al.},\ }\bibfield  {title} {\bibinfo
  {title} {The {{Sloan Lens ACS Survey}}. {{XIII}}. {{Discovery}} of 40 {{New
  Galaxy-scale Strong Lenses}}{${_\ast}$}},\ }\href
  {https://doi.org/10.3847/1538-4357/aa9794} {\bibfield  {journal} {\bibinfo
  {journal} {The Astrophysical Journal}\ }\textbf {\bibinfo {volume} {851}},\
  \bibinfo {pages} {48} (\bibinfo {year} {2017})}\BibitemShut {NoStop}%
\bibitem [{\citenamefont {Brownstein}\ \emph {et~al.}(2012)\citenamefont
  {Brownstein}, \citenamefont {Bolton}, \citenamefont {Schlegel}, \citenamefont
  {Eisenstein}, \citenamefont {Kochanek}, \citenamefont {Connolly},
  \citenamefont {Maraston}, \citenamefont {Pandey}, \citenamefont {Seitz} \emph
  {et~al.}}]{brownsteinBOSSEmissionLineLens2012}%
  \BibitemOpen
  \bibfield  {author} {\bibinfo {author} {\bibfnamefont {J.~R.}\ \bibnamefont
  {Brownstein}}, \bibinfo {author} {\bibfnamefont {A.~S.}\ \bibnamefont
  {Bolton}}, \bibinfo {author} {\bibfnamefont {D.~J.}\ \bibnamefont
  {Schlegel}}, \bibinfo {author} {\bibfnamefont {D.~J.}\ \bibnamefont
  {Eisenstein}}, \bibinfo {author} {\bibfnamefont {C.~S.}\ \bibnamefont
  {Kochanek}}, \bibinfo {author} {\bibfnamefont {N.}~\bibnamefont {Connolly}},
  \bibinfo {author} {\bibfnamefont {C.}~\bibnamefont {Maraston}}, \bibinfo
  {author} {\bibfnamefont {P.}~\bibnamefont {Pandey}}, \bibinfo {author}
  {\bibfnamefont {S.}~\bibnamefont {Seitz}}, \emph {et~al.},\ }\bibfield
  {title} {\bibinfo {title} {The {{BOSS Emission-Line Lens Survey}}
  ({{BELLS}}). {{I}}. {{A Large Spectroscopically Selected Sample}} of {{Lens
  Galaxies}} at {{Redshift}} \textasciitilde 0.5},\ }\href
  {https://doi.org/10.1088/0004-637X/744/1/41} {\bibfield  {journal} {\bibinfo
  {journal} {The Astrophysical Journal}\ }\textbf {\bibinfo {volume} {744}},\
  \bibinfo {pages} {41} (\bibinfo {year} {2012})}\BibitemShut {NoStop}%
\bibitem [{\citenamefont {Shu}\ \emph {et~al.}(2016)\citenamefont {Shu},
  \citenamefont {Bolton}, \citenamefont {Mao}, \citenamefont {Kochanek},
  \citenamefont {{P{\'e}rez-Fournon}}, \citenamefont {Oguri}, \citenamefont
  {{Montero-Dorta}}, \citenamefont {Cornachione}, \citenamefont
  {{Marques-Chaves}}, \citenamefont {Zheng} \emph
  {et~al.}}]{shuBOSSEmissionlineLens2016}%
  \BibitemOpen
  \bibfield  {author} {\bibinfo {author} {\bibfnamefont {Y.}~\bibnamefont
  {Shu}}, \bibinfo {author} {\bibfnamefont {A.~S.}\ \bibnamefont {Bolton}},
  \bibinfo {author} {\bibfnamefont {S.}~\bibnamefont {Mao}}, \bibinfo {author}
  {\bibfnamefont {C.~S.}\ \bibnamefont {Kochanek}}, \bibinfo {author}
  {\bibfnamefont {I.}~\bibnamefont {{P{\'e}rez-Fournon}}}, \bibinfo {author}
  {\bibfnamefont {M.}~\bibnamefont {Oguri}}, \bibinfo {author} {\bibfnamefont
  {A.~D.}\ \bibnamefont {{Montero-Dorta}}}, \bibinfo {author} {\bibfnamefont
  {M.~A.}\ \bibnamefont {Cornachione}}, \bibinfo {author} {\bibfnamefont
  {R.}~\bibnamefont {{Marques-Chaves}}}, \bibinfo {author} {\bibfnamefont
  {Z.}~\bibnamefont {Zheng}}, \emph {et~al.},\ }\bibfield  {title} {\bibinfo
  {title} {The {{BOSS Emission-line Lens Survey}}. {{IV}}. {{Smooth Lens
  Models}} for the {{BELLS GALLERY Sample}}},\ }\href
  {https://doi.org/10.3847/1538-4357/833/2/264} {\bibfield  {journal} {\bibinfo
   {journal} {The Astrophysical Journal}\ }\textbf {\bibinfo {volume} {833}},\
  \bibinfo {pages} {264} (\bibinfo {year} {2016})}\BibitemShut {NoStop}%
\bibitem [{\citenamefont {Cappellari}\ \emph {et~al.}(2006)\citenamefont
  {Cappellari}, \citenamefont {Bacon}, \citenamefont {Bureau}, \citenamefont
  {Damen}, \citenamefont {Davies}, \citenamefont {{de Zeeuw}}, \citenamefont
  {Emsellem}, \citenamefont {{Falc{\'o}n-Barroso}}, \citenamefont
  {Krajnovi{\'c}} \emph {et~al.}}]{cappellariSAURONProjectIV2006}%
  \BibitemOpen
  \bibfield  {author} {\bibinfo {author} {\bibfnamefont {M.}~\bibnamefont
  {Cappellari}}, \bibinfo {author} {\bibfnamefont {R.}~\bibnamefont {Bacon}},
  \bibinfo {author} {\bibfnamefont {M.}~\bibnamefont {Bureau}}, \bibinfo
  {author} {\bibfnamefont {M.~C.}\ \bibnamefont {Damen}}, \bibinfo {author}
  {\bibfnamefont {R.~L.}\ \bibnamefont {Davies}}, \bibinfo {author}
  {\bibfnamefont {P.~T.}\ \bibnamefont {{de Zeeuw}}}, \bibinfo {author}
  {\bibfnamefont {E.}~\bibnamefont {Emsellem}}, \bibinfo {author}
  {\bibfnamefont {J.}~\bibnamefont {{Falc{\'o}n-Barroso}}}, \bibinfo {author}
  {\bibfnamefont {D.}~\bibnamefont {Krajnovi{\'c}}}, \emph {et~al.},\
  }\bibfield  {title} {\bibinfo {title} {The {{SAURON}} project - {{IV}}.
  {{The}} mass-to-light ratio, the virial mass estimator and the {{Fundamental
  Plane}} of elliptical and lenticular galaxies},\ }\href
  {https://doi.org/10.1111/j.1365-2966.2005.09981.x} {\bibfield  {journal}
  {\bibinfo  {journal} {Monthly Notices of the Royal Astronomical Society}\
  }\textbf {\bibinfo {volume} {366}},\ \bibinfo {pages} {1126} (\bibinfo {year}
  {2006})}\BibitemShut {NoStop}%
\bibitem [{\citenamefont {Jiang}\ and\ \citenamefont
  {Kochanek}(2007)}]{jiangBaryonFractionsMasstoLight2007}%
  \BibitemOpen
  \bibfield  {author} {\bibinfo {author} {\bibfnamefont {G.}~\bibnamefont
  {Jiang}}\ and\ \bibinfo {author} {\bibfnamefont {C.~S.}\ \bibnamefont
  {Kochanek}},\ }\bibfield  {title} {\bibinfo {title} {The {{Baryon Fractions}}
  and {{Mass-to-Light Ratios}} of {{Early-Type Galaxies}}},\ }\href
  {https://doi.org/10.1086/522580} {\bibfield  {journal} {\bibinfo  {journal}
  {The Astrophysical Journal}\ }\textbf {\bibinfo {volume} {671}},\ \bibinfo
  {pages} {1568} (\bibinfo {year} {2007})}\BibitemShut {NoStop}%
\bibitem [{\citenamefont {{Foreman-Mackey}}\ \emph {et~al.}(2013)\citenamefont
  {{Foreman-Mackey}}, \citenamefont {Hogg}, \citenamefont {Lang},\ and\
  \citenamefont {Goodman}}]{foreman-mackeyEmceeMCMCHammer2013}%
  \BibitemOpen
  \bibfield  {author} {\bibinfo {author} {\bibfnamefont {D.}~\bibnamefont
  {{Foreman-Mackey}}}, \bibinfo {author} {\bibfnamefont {D.~W.}\ \bibnamefont
  {Hogg}}, \bibinfo {author} {\bibfnamefont {D.}~\bibnamefont {Lang}},\ and\
  \bibinfo {author} {\bibfnamefont {J.}~\bibnamefont {Goodman}},\ }\bibfield
  {title} {\bibinfo {title} {Emcee: {{The MCMC Hammer}}},\ }\href
  {https://doi.org/10.1086/670067} {\bibfield  {journal} {\bibinfo  {journal}
  {Publications of the Astronomical Society of the Pacific}\ }\textbf {\bibinfo
  {volume} {125}},\ \bibinfo {pages} {306} (\bibinfo {year} {2013})},\ \Eprint
  {https://arxiv.org/abs/1202.3665} {arxiv:1202.3665 [astro-ph,
  physics:physics, stat]} \BibitemShut {NoStop}%
\bibitem [{\citenamefont {Collaboration}\ \emph {et~al.}(2020)\citenamefont
  {Collaboration}, \citenamefont {Aghanim}, \citenamefont {Akrami},
  \citenamefont {Ashdown}, \citenamefont {Aumont}, \citenamefont {Baccigalupi},
  \citenamefont {Ballardini}, \citenamefont {Banday}, \citenamefont {Barreiro}
  \emph {et~al.}}]{planckcollaborationPlanck2018Results2020}%
  \BibitemOpen
  \bibfield  {author} {\bibinfo {author} {\bibfnamefont {P.}~\bibnamefont
  {Collaboration}}, \bibinfo {author} {\bibfnamefont {N.}~\bibnamefont
  {Aghanim}}, \bibinfo {author} {\bibfnamefont {Y.}~\bibnamefont {Akrami}},
  \bibinfo {author} {\bibfnamefont {M.}~\bibnamefont {Ashdown}}, \bibinfo
  {author} {\bibfnamefont {J.}~\bibnamefont {Aumont}}, \bibinfo {author}
  {\bibfnamefont {C.}~\bibnamefont {Baccigalupi}}, \bibinfo {author}
  {\bibfnamefont {M.}~\bibnamefont {Ballardini}}, \bibinfo {author}
  {\bibfnamefont {A.~J.}\ \bibnamefont {Banday}}, \bibinfo {author}
  {\bibfnamefont {R.~B.}\ \bibnamefont {Barreiro}}, \emph {et~al.},\ }\bibfield
   {title} {\bibinfo {title} {Planck 2018 results. {{VI}}. {{Cosmological}}
  parameters},\ }\href {https://doi.org/10.1051/0004-6361/201833910} {\bibfield
   {journal} {\bibinfo  {journal} {Astronomy \& Astrophysics}\ }\textbf
  {\bibinfo {volume} {641}},\ \bibinfo {pages} {A6} (\bibinfo {year} {2020})},\
  \Eprint {https://arxiv.org/abs/1807.06209} {arxiv:1807.06209 [astro-ph]}
  \BibitemShut {NoStop}%
\bibitem [{\citenamefont {Gerhard}\ \emph {et~al.}(2001)\citenamefont
  {Gerhard}, \citenamefont {Kronawitter}, \citenamefont {Saglia},\ and\
  \citenamefont {Bender}}]{gerhardDynamicalFamilyProperties2001}%
  \BibitemOpen
  \bibfield  {author} {\bibinfo {author} {\bibfnamefont {O.}~\bibnamefont
  {Gerhard}}, \bibinfo {author} {\bibfnamefont {A.}~\bibnamefont
  {Kronawitter}}, \bibinfo {author} {\bibfnamefont {R.~P.}\ \bibnamefont
  {Saglia}},\ and\ \bibinfo {author} {\bibfnamefont {R.}~\bibnamefont
  {Bender}},\ }\bibfield  {title} {\bibinfo {title} {Dynamical family
  properties and dark halo scaling relations of giant elliptical galaxies},\
  }\href {https://doi.org/10.1086/319940} {\bibfield  {journal} {\bibinfo
  {journal} {The Astronomical Journal}\ }\textbf {\bibinfo {volume} {121}},\
  \bibinfo {pages} {1936} (\bibinfo {year} {2001})},\ \Eprint
  {https://arxiv.org/abs/astro-ph/0012381} {arxiv:astro-ph/0012381}
  \BibitemShut {NoStop}%
\bibitem [{\citenamefont {Zhu}\ \emph {et~al.}(2023{\natexlab{a}})\citenamefont
  {Zhu}, \citenamefont {Lu}, \citenamefont {Cappellari}, \citenamefont {Li},
  \citenamefont {Mao},\ and\ \citenamefont
  {Gao}}]{zhuMaNGADynPopQualityassessed2023}%
  \BibitemOpen
  \bibfield  {author} {\bibinfo {author} {\bibfnamefont {K.}~\bibnamefont
  {Zhu}}, \bibinfo {author} {\bibfnamefont {S.}~\bibnamefont {Lu}}, \bibinfo
  {author} {\bibfnamefont {M.}~\bibnamefont {Cappellari}}, \bibinfo {author}
  {\bibfnamefont {R.}~\bibnamefont {Li}}, \bibinfo {author} {\bibfnamefont
  {S.}~\bibnamefont {Mao}},\ and\ \bibinfo {author} {\bibfnamefont
  {L.}~\bibnamefont {Gao}},\ }\href {https://doi.org/10.48550/arXiv.2304.11711}
  {\bibinfo {title} {{{MaNGA DynPop}} -- {{I}}. {{Quality-assessed}} stellar
  dynamical modelling from integral-field spectroscopy of {{10K}} nearby
  galaxies: A catalogue of masses, mass-to-light ratios, density profiles and
  dark matter}} (\bibinfo {year} {2023}{\natexlab{a}}),\ \Eprint
  {https://arxiv.org/abs/2304.11711} {arxiv:2304.11711 [astro-ph]} \BibitemShut
  {NoStop}%
\bibitem [{\citenamefont {Zhu}\ \emph {et~al.}(2023{\natexlab{b}})\citenamefont
  {Zhu}, \citenamefont {Lu}, \citenamefont {Cappellari}, \citenamefont {Li},
  \citenamefont {Mao},\ and\ \citenamefont {Gao}}]{zhuMaNGADynPopIII2023}%
  \BibitemOpen
  \bibfield  {author} {\bibinfo {author} {\bibfnamefont {K.}~\bibnamefont
  {Zhu}}, \bibinfo {author} {\bibfnamefont {S.}~\bibnamefont {Lu}}, \bibinfo
  {author} {\bibfnamefont {M.}~\bibnamefont {Cappellari}}, \bibinfo {author}
  {\bibfnamefont {R.}~\bibnamefont {Li}}, \bibinfo {author} {\bibfnamefont
  {S.}~\bibnamefont {Mao}},\ and\ \bibinfo {author} {\bibfnamefont
  {L.}~\bibnamefont {Gao}},\ }\href {https://doi.org/10.48550/arXiv.2304.11714}
  {\bibinfo {title} {{{MaNGA DynPop}} -- {{III}}. {{Accurate}} stellar dynamics
  vs. stellar population relations in 6000 early-type and spiral galaxies:
  Fundamental plane, mass-to-light ratios, total density slopes, and dark
  matter fractions}} (\bibinfo {year} {2023}{\natexlab{b}}),\ \Eprint
  {https://arxiv.org/abs/2304.11714} {arxiv:2304.11714 [astro-ph]} \BibitemShut
  {NoStop}%
\bibitem [{\citenamefont {Akaike}(1974)}]{akaikeNewLookStatistical1974}%
  \BibitemOpen
  \bibfield  {author} {\bibinfo {author} {\bibfnamefont {H.}~\bibnamefont
  {Akaike}},\ }\bibfield  {title} {\bibinfo {title} {A new look at the
  statistical model identification},\ }\href
  {https://doi.org/10.1109/TAC.1974.1100705} {\bibfield  {journal} {\bibinfo
  {journal} {IEEE Transactions on Automatic Control}\ }\textbf {\bibinfo
  {volume} {19}},\ \bibinfo {pages} {716} (\bibinfo {year} {1974})}\BibitemShut
  {NoStop}%
\bibitem [{\citenamefont
  {Schwarz}(1978)}]{schwarzEstimatingDimensionModel1978}%
  \BibitemOpen
  \bibfield  {author} {\bibinfo {author} {\bibfnamefont {G.}~\bibnamefont
  {Schwarz}},\ }\bibfield  {title} {\bibinfo {title} {Estimating the
  {{Dimension}} of a {{Model}}},\ }\href@noop {} {\bibfield  {journal}
  {\bibinfo  {journal} {The Annals of Statistics}\ }\textbf {\bibinfo {volume}
  {6}},\ \bibinfo {pages} {461} (\bibinfo {year} {1978})},\ \Eprint
  {https://arxiv.org/abs/2958889} {2958889} \BibitemShut {NoStop}%
\bibitem [{\citenamefont {Li}\ \emph {et~al.}(2018)\citenamefont {Li},
  \citenamefont {Shu},\ and\ \citenamefont
  {Wang}}]{liStronglensingMeasurementTotalmassdensity2018}%
  \BibitemOpen
  \bibfield  {author} {\bibinfo {author} {\bibfnamefont {R.}~\bibnamefont
  {Li}}, \bibinfo {author} {\bibfnamefont {Y.}~\bibnamefont {Shu}},\ and\
  \bibinfo {author} {\bibfnamefont {J.}~\bibnamefont {Wang}},\ }\bibfield
  {title} {\bibinfo {title} {Strong-lensing measurement of the
  total-mass-density profile out to three effective radii for z {$\sim$} 0.5
  early-type galaxies},\ }\href {https://doi.org/10.1093/mnras/sty1813}
  {\bibfield  {journal} {\bibinfo  {journal} {Monthly Notices of the Royal
  Astronomical Society}\ }\textbf {\bibinfo {volume} {480}},\ \bibinfo {pages}
  {431} (\bibinfo {year} {2018})}\BibitemShut {NoStop}%
\bibitem [{\citenamefont {Birrer}\ and\ \citenamefont
  {Amara}(2018)}]{birrerLenstronomyMultipurposeGravitational2018}%
  \BibitemOpen
  \bibfield  {author} {\bibinfo {author} {\bibfnamefont {S.}~\bibnamefont
  {Birrer}}\ and\ \bibinfo {author} {\bibfnamefont {A.}~\bibnamefont {Amara}},\
  }\bibfield  {title} {\bibinfo {title} {Lenstronomy: Multi-purpose
  gravitational lens modelling software package},\ }\href
  {https://doi.org/10.1016/j.dark.2018.11.002} {\bibfield  {journal} {\bibinfo
  {journal} {Physics of the Dark Universe}\ }\textbf {\bibinfo {volume} {22}},\
  \bibinfo {pages} {189} (\bibinfo {year} {2018})},\ \Eprint
  {https://arxiv.org/abs/1803.09746} {arxiv:1803.09746 [astro-ph]} \BibitemShut
  {NoStop}%
\end{thebibliography}%

\end{document}